\documentclass[12pt,a4paper]{article}
\usepackage{amsmath}
\usepackage{amssymb}
\usepackage{hyperref}
\usepackage{setspace}
\usepackage{graphicx} 
\usepackage{cite}

\newcommand{\malpha}{m_\alpha}
\newcommand{\mbeta}{m_\beta}

\begin{document}

\date{}
\title{{\Large \textbf{Momentum relaxation in holographic massive gravity}}}
\author{{\normalsize Richard A. Davison} \vspace{0.5cm} \\ {\normalsize \textit{Instituut-Lorentz for Theoretical Physics}} \\{\normalsize\textit{Niels Bohrweg 2, Leiden NL-2333 CA, The Netherlands}} \vspace{0.1cm} \\ {\normalsize davison@lorentz.leidenuniv.nl}}

\maketitle

\thispagestyle{empty}

\onehalfspacing
\begin{abstract}
We study the effects of momentum relaxation on observables in a recently proposed holographic model in which the conservation of momentum in the field theory is broken by the presence of a bulk graviton mass. In the hydrodynamic limit, we show that these effects can be incorporated by a simple modification of the energy-momentum conservation equation to account for the dissipation of momentum over a single characteristic timescale. We compute this timescale as a function of the graviton mass terms and identify the previously known ``wall of stability'' as the point at which this relaxation timescale becomes negative. We also calculate analytically the zero temperature AC conductivity at low frequencies. In the limit of a small graviton mass this reduces to the simple Drude form, and we compute the corrections to this which are important for larger masses. Finally, we undertake a preliminary investigation of the stability of the zero temperature black brane solution of this model, and rule out spatially modulated instabilities of a certain kind.

\end{abstract}

\clearpage
\tableofcontents

\section{Introduction}
\label{sec:Introduction}

Real materials, which are composed of electrons, atoms and so on, do not possess spatial translational invariance, and so momentum in these systems is not conserved. While the breaking of this symmetry -- for example by the presence of a lattice or of impurities -- is unimportant for some observables, it can give rise to important, qualitative effects in others. One simple example of such an observable is the conductivity in a system with a non-zero density of a conserved charge. Under an applied field, the charge carriers of this system will accelerate indefinitely if there is no way for them to dissipate momentum, leading to an infinite DC conductivity. If, however, spatial translational invariance is broken and therefore momentum dissipation is possible, the DC conductivity is finite and the delta function peak at the origin of the AC conductivity spreads out.

If holographic theories are to have a chance of explaining the experimentally determined behaviour of these kinds of observables, then it is important that they can incorporate the dissipation of momentum. There have been numerous studies of the transport properties of holographic theories which dissipate momentum by a variety of methods, such as by explicitly breaking the translational symmetry of the field theory state \cite{Horowitz:2012ky,Horowitz:2012gs,Horowitz:2013jaa,Hartnoll:2012rj,Donos:2012js,Flauger:2010tv,Aperis:2010cd,Iizuka:2012dk,Hutasoit:2012ib,Ooguri:2010kt,Kachru:2009xf,Kachru:2010dk,Liu:2012tr}, by including a parametrically large amount of neutral matter \cite{Karch:2007pd,Hartnoll:2009ns,Faulkner:2010da}, or by coupling to impurities \cite{Hartnoll:2007ih,Hartnoll:2008hs}. The recent work \cite{Mahajan:2013cja} addresses some of the general effects of momentum relaxation without reference to a specific underlying mechanism causing it, and in \cite{Sonner:2013aua} it was recently shown how one obtains a finite DC conductivity for current flows of a certain kind in translationally-invariant two-layer systems.

A holographic model was proposed recently in which, by giving a mass to the graviton, momentum is no longer conserved \cite{Vegh:2013sk}. This mass breaks the diffeomorphism invariance of the gravitational theory and therefore, via the holographic dictionary, it violates the conservation of energy-momentum in the dual field theory. One reason that this approach is attractive is that it is relatively simple from a practical point of view -- for example, the relevant black brane solutions are known analytically. This simplicity makes it an excellent toy model for studying the properties of holographic states of matter without momentum conservation. Theories of massive gravity are liable to be inconsistent (see, for example, \cite{Boulware:1973my,Gruzinov:2011sq,deRham:2011pt,Deser:2012qx,Deser:2013uy,Deser:2013eua} and references therein) and we do not know the microscopic details of the dual field theory (if one exists). As emphasised in \cite{Vegh:2013sk}, one could heuristically view this model as an effective theory arising from coarse-graining over the microscopic details of a bulk lattice or of impurities such that at long distances their only effect is to introduce effective mass terms for the graviton. We will address this comparison further when we present our results.

In this paper, we investigate in detail the effects of the non-conservation of momentum upon observable properties of the field theory state dual to the massive gravity solution of \cite{Vegh:2013sk}. Firstly, we outline how the extra bulk degrees of freedom produced by including a non-zero graviton mass result in the violation of the usual field theory Ward identities arising from momentum conservation. We then study this field theory in the hydrodynamic limit -- that is, when the temperature $T$ is much larger than the frequency $\omega$ and momentum $k$ of any excitation, and much larger than the rate at which momentum is dissipated. We propose that in this fluid-like limit, the low energy dynamics of the theory are governed by a modified conservation law for energy-momentum $T^{ab}$ such that for small perturbations around the equilibrium state where the fluid is at rest
\begin{equation}
\partial_a T^{at}=0, \;\;\;\;\;\;\;\;\;\;\;\;\;\partial_aT^{ai}=-\left(\epsilon+P\right)\tau_\text{rel.}^{-1}u^i=-\tau_\text{rel.}^{-1}T^{ti},
\end{equation}
where $\epsilon$, $P$ and $u^i$ are the energy density, pressure and velocity of the near-equilibrium field theory state, and the constant $\tau_\text{rel.}$ is the characteristic timescale of momentum relaxation in the theory. We verify this proposal by comparing the low-energy transverse excitations of this modified hydrodynamics to those computed from the massive gravity theory, and determine that the hydrodynamic momentum relaxation timescale is given by
\begin{equation}
\label{eq:IntroductionRelaxationTimeExpression}
\tau_\text{rel.}^{-1}=\frac{s}{2\pi\left(\epsilon+P\right)}\left(m_\alpha^2+m_\beta^2\right),
\end{equation} 
where $s$ is the entropy density of the field theory and $m_\alpha^2$ and $m_\beta^2$ (which, a priori, may be negative) are the two independent mass terms for the graviton. The condition $\tau_\text{rel.}=0$ is equivalent to the ``wall of stability'' found in \cite{Vegh:2013sk}. We can therefore give a physical meaning to this instability: the state is unstable when $\tau_\text{rel.}<0$ because it \textit{absorbs} momentum at a constant rate, rather than dissipating it, and thus small perturbations of the state will grow exponentially in time.

When the chemical potential $\mu$ of the theory is non-zero, the relaxation of momentum affects the dynamics of the conserved U(1) current $J^a$, and this is the second topic that we study. We compute analytically the low frequency AC conductivity $\sigma\left(\omega\right)$ of the state when $T=0,\mu\ne0$ and $m_\alpha=0$ and find that
\begin{equation}
\label{eq:IntroductionFullConductivityExpression}
\sigma\left(\omega\right)=\frac{\sigma_{\text{DC}}+\ldots}{1+\kappa_1i\omega\log\left(\frac{\omega r_0}{6-m_\beta^2r_0^2}\right)+\kappa_2\omega+\kappa_3i\omega+\ldots},
\end{equation}
where $\sigma_\text{DC}$ is the $m_\beta$-dependent DC conductivity given in equation (\ref{eq:OurDCConductivityResult}), $\kappa_i$ are $\omega$-independent and $m_\beta$-dependent quantities given in equation (\ref{eq:kappadefinitions}) and where the ellipses denote higher order terms in $\omega$. In the limit of small graviton mass $m_\beta^2/\mu^2\ll\omega/\mu\ll1$, where momentum conservation is violated in a minor way, the conductivity is equivalent to that predicted by the simple Drude model
\begin{equation}
\sigma\left(\omega\right)=\frac{\sigma_\text{DC}}{1-i\omega\tau_\text{rel.}},
\end{equation}
where $\tau_\text{rel.}$ is in fact given by the naive $T=0$ limit of the hydrodynamic formula (\ref{eq:IntroductionRelaxationTimeExpression}). However, the full expression (\ref{eq:IntroductionFullConductivityExpression}) includes corrections to the Drude model. The inclusion of these corrections results in a transfer of spectral weight from the Drude peak to higher frequencies, and a reduction in the phase of $\sigma$ from the Drude value. Over a range of intermediate frequencies it was shown numerically in \cite{Vegh:2013sk} that the conductivity exhibits an approximate scaling law which, with the appropriate choice of $m_\beta^2$, is similar to that seen in holographic lattice models \cite{Horowitz:2012ky,Horowitz:2012gs} and is also reminiscent of that measured in the normal phase of some high-$T_c$ superconductors \cite{experimentalpaper1}. Unfortunately, the expression (\ref{eq:IntroductionFullConductivityExpression}) for the conductivity is perturbative in $\omega$ and contains no hints of this scaling behaviour seen at (relatively) high $\omega$.

Finally, we make a preliminary investigation of the possible existence of instabilities of the $T=0$ state to a spatially modulated phase by computing the $k$-dependent masses of the bulk field excitations transverse to the momentum flow in the near-horizon AdS$_2$ geometry when $m_\alpha=0$. We find that all of these masses satisfy the Breitenlohner-Freedman bound for all values of $k$ and thus an instability of this specific kind is not present.

The outline of the remainder of the paper is as follows. In section \ref{sec:BackgroundSolution}, we briefly review the action and relevant black brane solution of massive gravity introduced in \cite{Vegh:2013sk} and in section \ref{sec:FluctuationEquationsAndAction} we present the equations of motion and on-shell action of linearised transverse fluctuations in this theory, emphasising how the breaking of bulk diffeomorphism invariance can be explicitly seen to violate the field theory Ward identities due to momentum conservation. In section \ref{sec:ModifiedHydrodynamicsAndComparison} we outline how hydrodynamics should be modified to account for the non-conservation of momentum, and compute the timescale of momentum relaxation by a study of the low energy transverse excitations of the dual field theory. Section \ref{sec:ZeroTemperatureConductivityAnalysis} contains an analytic derivation of the low frequency conductivity of the zero temperature field theory state. In section \ref{sec:PreliminaryStabilityStudy} we begin an exploration of the stability of the solution of interest before finishing in section \ref{sec:DiscussionSection} with a summary of our results and some suggestions for future research. The appendix contains some simplifications of the fluctuation equations in the limit $m_\alpha=0$, including a proof of the decoupling of the two `master fields' when $k=0$.

\section{Massive gravity and its equilibrium solution}
\label{sec:BackgroundSolution}

The non-linear theory of massive gravity that we will investigate couples the metric tensor $g_{\mu\nu}$ to a fixed reference metric $f_{\mu\nu}$, giving a mass to $g_{\mu\nu}$ and breaking diffeomorphism invariance. With the usual holographic motivation, we also include a negative cosmological constant and a minimally coupled Maxwell field $A_\mu$ in the action \cite{Vegh:2013sk}
\begin{equation}
\label{eq:MassiveGravityAction}
S=\frac{1}{2\kappa_4^2}\int d^4x\sqrt{-g}\left(\mathcal{R}+\frac{6}{L^2}-\frac{L^2}{4}F_{\mu\nu}F^{\mu\nu}+m^2\left\{\alpha\left[\mathcal{K}\right]+\beta\left(\left[\mathcal{K}\right]^2-\left[\mathcal{K}^2\right]\right)\right\}\right),
\end{equation}
where $\alpha,\beta$ are arbitrary dimensionless coupling constants, square brackets denote the trace $\left[\mathcal{K}\right]=\mathcal{K}^\mu_\mu$, indices are raised and lowered with the dynamical metric $g_{\mu\nu}$ and $\mathcal{K}$ satisfies $\mathcal{K}^\mu_\alpha\mathcal{K}^\alpha_\nu=g^{\mu\alpha}f_{\alpha\nu}$ where the reference metric is chosen to be $f_{xx}=f_{yy}=F^2$ for a constant $F$, with all other components vanishing. Our co-ordinates are $\left(t,x,y,r\right)$ where $r$ is the holographic radial direction. This choice of reference metric clearly distinguishes the two spatial directions $x,y$ from the temporal and radial directions $t,r$. This choice breaks the symmetries associated with reparameterisations of the spatial $x,y$ co-ordinates and will result in the dissipation of momentum (but not energy) in the dual field theory. It was shown in \cite{Vegh:2013sk}, following \cite{Hassan:2011tf}, that the Boulware-Deser ghost \cite{Boulware:1973my} may be absent in this theory (see \cite{Hinterbichler:2011tt} for a pedagogical review of theoretical aspects of massive gravity). We will not comment further upon the theory's non-linear stability. Instead, we are primarily interested in whether any purported field theory dual of (\ref{eq:MassiveGravityAction}) is a sensible model for a condensed matter system which dissipates momentum. We will find, at the level of the two-point functions of the supposed field theory, that it is. We will henceforth assume the existence of a strongly-coupled field theory dual to (\ref{eq:MassiveGravityAction}), and use the usual holographic dictionary \cite{Gubser:1998bc,Witten:1998qj,Aharony:1999ti,Hartnoll:2009sz,McGreevy:2009xe} to compute its properties.

The action (\ref{eq:MassiveGravityAction}) admits the black brane solution \cite{Vegh:2013sk}
\begin{equation}
\begin{aligned}
\label{eq:backgroundsolution}
ds^2&=\frac{L^2}{r^2}\left(-f(r)dt^2+dx^2+dy^2+\frac{dr^2}{f(r)}\right),\;\;\;\;\;\;\;\;\;\;\;\;\;\;\; A_t\left(r\right)=\mu\left(1-\frac{r}{r_0}\right),\\
f(r)&=1-r_0^2m_\alpha^2\frac{r}{r_0}-r_0^2m_\beta^2\frac{r^2}{r_0^2}-\left[1-r_0^2m_\alpha^2-r_0^2m_\beta^2+\frac{1}{4}r_0^2\mu^2\right]\frac{r^3}{r_0^3}+\frac{1}{4}r_0^2\mu^2\frac{r^4}{r_0^4},
\end{aligned}
\end{equation}
where the radial co-ordinate $r$ takes values between $0$ (the boundary of the spacetime) and $r_0$, the location of the black brane horizon, and the constants $m_\alpha^2$ and $m_\beta^2$ (which may be either positive or negative) are related to the parameters of the action (\ref{eq:MassiveGravityAction}) via
\begin{equation}
m_\alpha^2=-\frac{\alpha FLm^2}{2r_0},\;\;\;\;\;\;\;\;\;\;\;\;\;\;\;\;\;\;\; m_\beta^2=-\beta F^2m^2.
\end{equation}
This solution is invariant under translations in the spatial field theory directions $x,y$ (as well as in the temporal direction). As we will show in the following section, it is the lack of diffeomorphism invariance of the linearised fluctuations that will lead to momentum dissipation in the dual field theory. The diffeomorphism-breaking terms $m_\alpha^2$ and $m_\beta^2$ appear as mass terms for the linearised metric perturbations. 

The temperature $T$ of the dual field theory state is given by
\begin{equation}
\label{eq:TemperatureDefinition}
T=\frac{1}{4\pi r_0}\left[3-2\left(r_0m_\alpha\right)^2-\left(r_0m_\beta\right)^2-\frac{1}{4}\left(r_0\mu\right)^2\right],
\end{equation}
and the chemical potential of the state is $\mu$. A non-zero $\mu$ results in a non-zero density of charge in the field theory state. Further thermodynamic properties of this solution were studied in \cite{Vegh:2013sk}.\footnote{Note added: A more careful study of the thermodynamics has subsequently been performed in \cite{Blake:2013bqa}.} This state has four independent dimensionful scales: $r_0,\mu,m_\alpha^2,m_\beta^2$. In the field theory, it is easier to think in terms of the more physically motivated scales $T,\mu,m_\alpha^2,m_\beta^2$, where we will later give some physical meaning to $m_\alpha^2$ and $m_\beta^2$ in the field theory in terms of the momentum relaxation timescale. Near the horizon, the zero temperature black brane geometry has the form AdS$_2\times\mathbb{R}^2$, which we will utilise in detail in section \ref{sec:ZeroTemperatureConductivityAnalysis} when computing the AC conductivity.

\section{Linearised fluctuations and Ward identities}
\label{sec:FluctuationEquationsAndAction}

To determine the transport properties of the dual field theory, we must study fluctuations of the bulk fields $g_{\mu\nu}$ and $A_{\mu}$ around the background (\ref{eq:backgroundsolution}), which is translationally invariant in the $\left(t,x,y\right)$ directions. It is simplest to work in Fourier space and thus we write
\begin{equation}
\begin{aligned}
&g_{\mu\nu}\left(r\right)\rightarrow g_{\mu\nu}\left(r\right)+\int\frac{d\omega dk}{\left(2\pi\right)^2}e^{-i\omega t+ikx}h_{\mu\nu}\left(r,\omega,k\right),\\
&A_{\mu}\left(r\right)\rightarrow A_{\mu}\left(r\right)+\int\frac{d\omega dk}{\left(2\pi\right)^2}e^{-i\omega t+ikx}a_{\mu}\left(r,\omega,k\right).
\end{aligned}
\end{equation}
Note that the Fourier transform defined here has the opposite sign in the exponent than the transform defined in \cite{Vegh:2013sk}. For our purposes (which are to compute two-point Green's functions), it is sufficient to study the linearised fluctuations. As usual, the fluctuations can be classified as either odd or even under the transformation $y\rightarrow-y$. As both the action (\ref{eq:MassiveGravityAction}) and the background solution (\ref{eq:backgroundsolution}) are invariant under this transformation, the even fluctuations ($h_{rt}, h_{rx}, h_{xt}, h_{tt}, h_{xx}, h_{yy}, h_{rr}, a_{r}, a_{x}, a_{t}$) decouple from the odd ones ($h_{yx}, h_{yt}, h_{yr}, a_{y}$) at linearised order. We will refer to the fields which are odd/even under this transformation as transverse/longitudinal (with respect to the direction of $k$). 

As in the usual gauge/gravity dictionary, we take $A^\mu$ to be dual to a conserved U(1) current $J^a$ in the dual field theory and $g_{\mu\nu}$ to be dual to the energy-momentum tensor $T^{ab}$, which we will shortly show is no longer conserved.

\subsection{Equations of motion with reduced gauge invariance}

For the usual case of a theory with diffeomorphism invariance, these linearised fluctuations possess a gauge symmetry under infinitesimal co-ordinate transformations $\delta_\xi x^{\mu}=\xi^\mu\left(x\right)$ which act as
\begin{equation}
\label{eq:infinitesimaldiffeos}
\delta_\xi h_{\mu\nu}=-\bar{\nabla}_\mu\xi_\nu-\bar{\nabla}_\nu\xi_\mu,\;\;\;\;\;\;\;\;\;\;\;\;\;\;\;\;\delta_\xi a_\mu=-\xi^\alpha\bar{\nabla}_\alpha A_\mu-A_{\alpha}\bar{\nabla}_\mu\xi^\alpha,
\end{equation}
where the bar denotes a covariant derivative with respect to the background metric (\ref{eq:backgroundsolution}). In addition, there is an invariance under infinitesimal U(1) gauge transformations of the form
\begin{equation}
\delta_\Lambda a_\mu=a_\mu-\partial_\mu\Lambda,\;\;\;\;\;\;\;\;\;\;\;\;\;\;\;\;\delta_\Lambda h_{\mu\nu}=0.
\end{equation}
These bulk gauge symmetries encode global symmetries of the dual field theory, as was noted during the original proposition of the AdS/CFT correspondence \cite{Maldacena:1997re}. At the level of linearised fluctuations, it is possible to make this encoding very explicit by working with certain gauge-invariant combinations of fluctuations. This then allows a clearer picture of the repercussions for the field theory of breaking the bulk diffeomorphism invariance.

From the action (\ref{eq:MassiveGravityAction}), the linearised equations of motion of the transverse fluctuations are
\begin{subequations}
\begin{align}
& \frac{d}{dr}\left[\frac{1}{r^2}\left({h^y_t}'+i\omega h^y_r\right)+A_t'a_y\right]-\frac{k}{r^2f}\left(kh^y_t+\omega h^y_x\right)-\frac{2}{r^2f}\left(\mbeta^2+\malpha^2\frac{r_0}{r}\right)h^y_t=0, \label{eq:HytEoM}\\
& \frac{d}{dr}\left[\frac{f}{r^2}\left({h^y_x}'-ikh^y_r\right)\right]+\frac{\omega}{r^2f}\left(\omega h^y_x+kh^y_t\right)-\frac{r_0\malpha^2}{r^3}h^y_x=0, \label{eq:HyxEoM}\\
& i\omega{h^y_t}'+ikf{h^y_x}'-\left[\omega^2-k^2f-2f\left(\mbeta^2+\malpha^2\frac{r_0}{r}\right)\right]h^y_r+i\omega r^2A_t'a_y=0, \label{eq:HyrEoM}\\
& \frac{d}{dr}\left[fa_y'\right]+A_t'\left({h^y_t}'+i\omega h^y_r\right)+\frac{1}{f}\left(\omega^2-k^2f\right)a_y=0, \label{eq:ayEoM}
\end{align}
\end{subequations}
where a prime denotes a derivative with respect to $r$ and indices are raised and lowered using the background metric (\ref{eq:backgroundsolution}). These equations are equivalent to those given in \cite{Vegh:2013sk} after setting $k=0$, changing the sign of $\omega$ (due to the opposite sign used in the definition of the Fourier transform) and rescaling $a_y$ by a factor of 2.

The main effect of the non-diffeomorphism-invariant terms in the action -- apart from changing the background function $f(r)$ appearing in the fluctuation equations -- is to produce explicit mass terms for the transverse components of $h_{\mu\nu}$ in the equations of motion above. That these mass terms break diffeomorphism invariance can be explicitly verified by noting that the equations of motion (\ref{eq:HytEoM}), (\ref{eq:HyxEoM}) and (\ref{eq:HyrEoM}) with $m^2_\alpha,m^2_\beta\ne0$ are not invariant under the infinitesimal diffeomorphisms (\ref{eq:infinitesimaldiffeos}), under which the transverse fields transform as
\begin{equation}
\label{eq:specificfielddiffeomorphismtransformations}
\delta_\xi h^y_r=-\frac{r^2}{L^2}\left(\xi_y'+\frac{2}{r}\xi_y\right),\;\;\;\;\;\;\;\;\delta_\xi h^y_t=\frac{i\omega r^2}{L^2}\xi_y,\;\;\;\;\;\;\;\;\delta_\xi h^y_x=-\frac{ikr^2}{L^2}\xi_y,\;\;\;\;\;\;\;\;\delta_\xi a_y=0.
\end{equation}
As a consequence, we see that diffeomorphisms with $\xi_y\ne0$ are no longer symmetries of the gravitational theory. 

As one might expect, this loss of a gauge symmetry results in the creation of an extra dynamical degree of freedom. To best parameterise this, let us firstly recall how to write the massless theory in an explicitly diffeomorphism-invariant form \cite{Kovtun:2005ev}. To do this, we wish to select a set of diffeomorphism-invariant bulk fields to study, rather than the gauge-dependent fundamental fluctuations $h^y_t$ etc. There is no unique way to select such a set from the bulk fields and their derivatives, but there is a natural choice to make. Fields $h_{\mu\nu}$ with both indices in the `field theory directions' $(t,x,y)$ are dual to components of the conserved energy-momentum tensor $T^{ab}$ of the strongly-coupled field theory, whereas the components of $h_{\mu\nu}$ with indices in the $r$-direction have no such direct field theory interpretation. The natural choice is therefore to study gauge-invariant combinations of bulk fields which do not involve $h_{\mu r}$. With this restriction, the only diffeomorphism-invariant combinations of the transverse fields are
\begin{equation}
\label{eq:diffinvariantcombinations}
Z_1=\frac{\omega}{k}h^y_x+h^y_t,\;\;\;\;\;\;\;\;\;\;\;\;\;\;\;\; Z_2=a_y,
\end{equation}
or any linear combination of them. There is a simple procedure to obtain the equations of motion for these gauge-invariant combinations -- one simply solves the `constraint' equation (\ref{eq:HyrEoM}) (in the massless limit) for the constraint field $h^y_r$, and then substitutes this solution into the remaining `dynamical' equations of motion (\ref{eq:HytEoM}), (\ref{eq:HyxEoM}) and (\ref{eq:ayEoM}) (in the massless limit). The resulting three equations are comprised of only two linearly independent equations which are the two coupled, dynamical (i.e. second order) differential equations for the gauge-invariant fields $Z_{1,2}$.

By repeating this procedure of substituting for the constraint field $h^y_r$ in the on-shell action for the fluctuations, one can write it in a manifestly gauge-invariant form
\begin{equation}
\begin{aligned}
S=\int_{r\rightarrow0}\frac{d\omega dk}{\left(2\pi\right)^2}\left[Z_i^{(0)}\left(-\omega,-k\right)G_{ij}\left(\omega,k\right)Z_j^{(0)}\left(\omega,k\right)+\ldots\right],
\end{aligned}
\end{equation}
where the superscript $(0)$ denotes the value of the field at the boundary, the ellipsis denotes contact terms (i.e. terms analytic in $\omega$ and $k$) and the explicit form of $G_{ij}\left(\omega,k\right)$ can be found in \cite{Davison:2013bxa}. With this formalism, it is clear that the the retarded Green's functions of the dual operators, extracted via the usual holographic procedure, obey (up to contact terms) \cite{Kovtun:2005ev,Herzog:2009xv}
\begin{equation}
\label{eq:wardidentities}
G^R_{T^{xy}T^{xy}}=\frac{\omega}{k}G^R_{T^{ty}T^{xy}}=\frac{\omega^2}{k^2}G^R_{T^{ty}T^{ty}},\;\;\;\;\;\;\;\text{etc.}
\end{equation}
These are simply the Ward identities due to conservation of (transverse) energy-momentum in the field theory: $\partial_aT^{ay}=0$. This explicitly gauge-invariant formalism therefore manifestly encodes the corresponding global symmetries of the dual field theory (at the level of the two-point functions).

We will now repeat this procedure when $m^2_\alpha,m^2_\beta\ne0$. As before, it is simple to solve the constraint equation (\ref{eq:HyrEoM}) for $h^y_r$ and substitute this solution into the remaining dynamical equations of motion (\ref{eq:HytEoM}), (\ref{eq:HyxEoM}) and (\ref{eq:ayEoM}). As before, we have now eliminated $h^y_r$ from our equations and will ascribe no meaning to this bulk field from the point of view of the dual field theory. In contrast to the massless case, the three resulting dynamical equations of motion are all linearly independent. They may be written
\begin{subequations}
\begin{align}
& \frac{d}{dr}\left[\frac{f}{r^2}\frac{\omega\left(k{h^y_t}'+\omega{h^y_x}'\right)-2f\left(m^2_\beta+m^2_\alpha\frac{r_0}{r}\right){h^y_x}'+\omega kr^2A_t'a_y}{\omega^2-k^2f-2f\left(m^2_\beta+m^2_\alpha\frac{r_0}{r}\right)}\right]+\frac{\omega}{r^2f}\left(\omega h^y_x+kh^y_t\right)-\frac{r_0m_\alpha^2}{r^3}h^y_x=0, \label{eq:SubstitutedEoM1}\\
& \frac{d}{dr}\left[fa_y'\right]-A_t'\frac{kf\left(k{h^y_t}'+\omega{h^y_x}'\right)+2f\left(m_\beta^2+m_\alpha^2\frac{r_0}{r}\right){h^y_t}'+\omega^2r^2A_t'a_y}{\omega^2-k^2f-2f\left(m_\beta^2+m_\alpha^2\frac{r_0}{r}\right)}+\frac{1}{f}\left(\omega^2-k^2f\right)a_y=0, \label{eq:SubstitutedEoM2}\\
& \frac{d}{dr}\left[\frac{f\left(m^2_\beta+m^2_\alpha\frac{r_0}{r}\right)\left(\omega{h^y_t}'+kf{h^y_x}'+\omega r^2A_t'a_y\right)}{r^2\left[\omega^2-k^2f-2f\left(m^2_\beta+m^2_\alpha\frac{r_0}{r}\right)\right]}\right]+\frac{\omega}{r^2f}\left(m^2_\beta+m^2_\alpha\frac{r_0}{r}\right)h^y_t+\frac{kr_0m^2_\alpha}{2r^3}h^y_x=0. \label{eq:SubstitutedEoM3}
\end{align}
\end{subequations}
Due to the breaking of diffeomorphism invariance, the fields $h^y_x$ and $h^y_t$ no longer appear in the previously-gauge-invariant combination $Z_1$ but are now independent dynamical degrees of freedom. This means that the two point functions of $T^{xy}$ and of $T^{ty}$ in the dual field theory are now independent and that the Ward identities (\ref{eq:wardidentities}) are no longer satisfied. This, of course, is a consequence of the fact that (transverse) energy-momentum is no longer conserved in the dual field theory: $\partial_aT^{ay}\ne0$. In total there are now three independent, dynamical transverse fields -- $h^y_t$, $h^y_x$, and $a_y$ -- which are gauge-invariant with respect to the unbroken gauge symmetries (diffeomorphisms in the $r,t$ directions and the U(1) gauge symmetry), which is one more than in the massless case as we previously anticipated. Note that in some situations (such as when computing the AC conductivity as we will do in section \ref{sec:ZeroTemperatureConductivityAnalysis}) it is more convenient to work with more involved bulk variables than the three fundamental fields themselves, but this does not affect the argument just outlined. 

Similar effects -- the generation of extra bulk dynamical degrees of freedom and the relaxation of field theory Ward identities due to conservation of energy-momentum -- will occur in the longitudinal sector of the theory.

\subsection{On-shell action}

Near the boundary of the spacetime (\ref{eq:backgroundsolution}), each field $h^y_t$, $h^y_x$, $a_y$ can be characterised by the two independent coefficients in its near-boundary power series expansion. Just as in the massless case, for $h^y_t$ and $h^y_x$ these are the coefficients of the terms of order $r^0$ and $r^3$ in said expansions, and for $a_y$ they are the coefficients of the terms of order $r^0$ and $r^1$. However, a non-zero value of $m_\alpha^2$ does make one qualitative difference to these near-boundary expansions: it produces terms in the near-boundary expansions of all three fields which are logarithmic in $r$, in addition to the usual integer powers. Such terms usually arise in bulk theories with an odd number of dimensions and are related to the conformal anomaly of the dual field theory \cite{Henningson:1998gx}, but they appear here in an even-dimensional theory. These logarithms will not be important in what follows. In fact, all of the interesting qualitative features of our results can be found even when $m_\alpha^2=0$.

To compute observables in the dual field theory, we require the on-shell gravitational action. After eliminating $h^y_r$ using the constraint equation (\ref{eq:HyrEoM}) and including the Gibbons-Hawking term in the action, this is given by
\begin{equation}
\begin{aligned}
\label{eq:onshellactionfullformula}
S=\frac{L^2}{2\kappa_4^2}\int_{r\rightarrow0}&\frac{d\omega dk}{\left(2\pi\right)^2}\Bigl\{\frac{f}{2r^2}\frac{1}{\left[\omega^2-k^2f-2f\left(m_\beta^2+m_\alpha^2\frac{r_0}{r}\right)\right]}\Bigl[-\left(\omega h^y_x+kh^y_t\right)\left(\omega{h^y_x}'+k{h^y_t}'\right)\\
&-2\left(m_\beta^2+m_\alpha^2\frac{r_0}{r}\right)\left(h^y_t{h^y_t}'-fh^y_x{h^y_x}'\right)\Bigr]-\frac{f}{2}a_ya_y'+\text{non-derivative terms}\Bigr\},
\end{aligned}
\end{equation}
where a prime denotes a derivative with respect to $r$ and the arguments of the first and second fluctuation in each pair are $\left(r,-\omega,-k\right)$ and $\left(r,\omega,k\right)$ respectively. As expected from the breaking of diffeomorphism invariance just outlined, the fields in the on-shell action no longer appear in the diffeomorphism-invariant combinations (\ref{eq:diffinvariantcombinations}).

The non-derivative terms, which we have not written explicitly, include the counterterms required to render the on-shell action finite. For the massless theory these counterterms are given, for example, in \cite{Edalati:2010hk}. We assume that it is possible to write an analogous set of counterterms in the case with non-zero mass such that the on-shell action is finite. These counterterms, and in fact all of the non-derivative terms in (\ref{eq:onshellactionfullformula}), will not affect any of the results we present because they contribute real contact terms (i.e. real terms analytic in $\omega,k$) to the retarded Greens functions $G^R_{\mathcal{O}\mathcal{O}}\left(\omega,k\right)$ of operators in the dual theory. Most of the quantities which we compute holographically -- poles of the Greens functions, and spectral functions -- are not affected by such terms. The only quantity we compute which is sensitive to such terms is the imaginary part of the conductivity $\sigma\left(\omega\right)$ which will receive non-zero contributions from any $\sim a_y^2$ counterterms. However, since the part of the on-shell action (\ref{eq:onshellactionfullformula}) dependent upon $a_y$ is already finite, we anticipate that there will be no counterterms of this form.

\section{``Hydrodynamics", momentum relaxation and the wall of stability}
\label{sec:ModifiedHydrodynamicsAndComparison}

A ubiquitous feature of translationally invariant media at non-zero temperatures is the applicability of hydrodynamics as an effective theory at small enough frequencies $\omega$ and momenta $k$. Hydrodynamics is a fluid-like limit based on the assumptions that the energy-momentum tensor and any global currents of a theory can be expressed in terms of a small number of slowly varying (with respect to a microscopic length scale $l_{\text{mfp}}$) macroscopic variables and their derivatives, and that the energy-momentum tensor and global currents are conserved \cite{LandauLifshitz,Kovtun:2012rj}. These assumptions should be valid for a state in local thermal equilibrium. For simplicity, let us consider the case of a (2+1)-dimensional conformal fluid at non-zero temperature and with no conserved charges. In this case, the constitutive relation expressing the energy-momentum tensor in terms of macroscopic variables and their derivatives is \cite{Baier:2007ix}
\begin{equation}
\label{eq:emtensorconstitutiverelation}
T^{ab}=\left(\epsilon+P\right)u^a u^b+Pg^{ab}-\eta\left(g^{ac}+u^a u^c\right)\left(g^{bd}+u^b u^d\right)\left(\partial_c u_d+\partial_d u_c -g_{cd}\partial_e u^e\right)+\ldots,
\end{equation}
where $\epsilon$ is the energy density in the fluid rest frame, $P$ is the fluid pressure, $u^a$ is the fluid three-velocity which obeys $u^au_a=-1$, $\eta$ is the shear viscosity, $g_{ab}$ is the Minkowski metric and the ellipsis denotes terms which are higher order in spacetime derivatives and therefore suppressed at small frequencies and momenta by powers of $\omega l_\text{mfp},kl_\text{mfp}\ll1$. 

Hydrodynamics predicts the existence of two long-lived collective excitations in such a theory: a sound mode due to longitudinal fluid flow (i.e. $u^i$ parallel to $k$, where $i$ labels the spatial directions of the field theory) and a shear diffusion mode due to transverse fluid flow ($u^i$ transverse to $k$). To show the existence of these modes, consider a field theory state which has been perturbed slightly from equilibrium such that
\begin{equation}
\label{eq:neareqmfluidflow}
u_a=\left(-1,\delta u_x,\delta u_y\right),\;\;\;\;\;\;\;\;\;\;\;\;\;\;\; \epsilon\rightarrow\epsilon+\delta\epsilon,\;\;\;\;\;\;\;\;\;\;\;\;\;\;\; P\rightarrow P+\delta P.
\end{equation}
Without loss of generality, we choose the momentum of the perturbation $k$ to flow in the $x$-direction so that $\delta u_a$ are functions of $x$ and $t$ only. The energy-momentum tensor of a fluid in this state is given by substituting this profile for $u_a$ into (\ref{eq:emtensorconstitutiverelation}). Demanding that $T^{ab}$ is conserved for this configuration, we find (after Fourier transforming and keeping terms which are linear in the amplitude of perturbation) that this is only possible for perturbations satisfying specific dispersion relations. For longitudinal fluid flow ($\delta u_y=0$), this is the dispersion relation of the sound mode and for transverse fluid flow ($\delta\epsilon=\delta P=\delta u_x=0$), this is the dispersion relation of the shear diffusion mode
\begin{equation}
\omega=-i\frac{\eta}{\epsilon+P}k^2+\ldots,
\end{equation}
where the ellipsis denotes higher order terms in $k$. Upon Fourier transforming to real space, this dispersion relation is that of a purely decaying mode (i.e. it has a vanishing propagation frequency) with decay rate $\Gamma=\eta k^2/\left(\epsilon+P\right)$ or equivalently with a lifetime $\tau=\left[\eta k^2/\left(\epsilon+P\right)\right]^{-1}$.

In the language of quantum field theory, these collective modes are realised as poles in the two-point functions of the longitudinal and transverse components of $T^{ab}$ respectively. These predictions of hydrodynamics (and many others) have been comprehensively verified for holographic theories with momentum conservation \cite{Policastro:2001yc,Policastro:2002se,Policastro:2002tn,Son:2007vk,Bhattacharyya:2008jc,Banerjee:2008th,Hubeny:2011hd}, and in particular they have been verified for the massless version of the theory \cite{Herzog:2002fn,Herzog:2003ke,Saremi:2006ep,Herzog:2007ij,Hartnoll:2007ai,Hartnoll:2007ip,Hartnoll:2007ih,Ge:2010yc,WitczakKrempa:2012gn,WitczakKrempa:2013ht} which we are studying. For holographic theories, a sufficient condition for the applicability of hydrodynamics is that we consider perturbations with $\omega,k\ll T$ (i.e. $l_\text{mfp.}\sim T^{-1}$).

\subsection{Modified hydrodynamics}
\label{sec:ModifiedHydrodynamicsSection}

After violating the conservation of momentum in our field theory by the inclusion of non-zero mass terms $m^2_\alpha,m^2_\beta\ne0$ in its gravitational dual, we no longer expect hydrodynamics to be the correct effective theory at low energies since energy-momentum is no longer conserved. In the following subsections, we will determine the effect of these non-zero mass terms on the shear diffusion mode of our field theory by computing the relevant Green's functions from its gravitational dual. Firstly though, we will propose a modification of hydrodynamics which replicates these effects. 

Our proposal is to consider a fluid with the same constitutive relation (\ref{eq:emtensorconstitutiverelation}) as previously but, for small perturbations around the equilibrium state where the fluid is at rest, to replace the conservation equation with
\begin{equation}
\label{eq:nonconservationproposal}
\partial_a T^{at}=0, \;\;\;\;\;\;\;\;\;\;\;\;\;\partial_aT^{ai}=-\left(\epsilon+P\right)\tau_\text{rel.}^{-1}u^i,
\end{equation}
where $\tau_\text{rel.}$ is a constant in spacetime. A similar modification was proposed in \cite{Hartnoll:2007ih} in the context of impurity scattering. What does this mean? Consider a spatially independent but time-dependent perturbation of the fluid. Then, according to equation (\ref{eq:nonconservationproposal}), the energy density of the fluid is constant in time $\partial _t T^{tt}=0$, but the momentum density of the fluid is now time-dependent: $\partial_tT^{ti}\ne0$. The fluid will lose momentum at a rate proportional to its velocity, with proportionality constant $\left(\epsilon+P\right)\tau_\text{rel.}^{-1}$. Specifically, for a spatially-independent near-equilibrium fluid flow of the form (\ref{eq:neareqmfluidflow}), $T^{ti}=\left(\epsilon+P\right)u^i$ and hence our modification of the conservation law corresponds to
\begin{equation}
\partial_t T^{tt}=0, \;\;\;\;\;\;\;\;\;\;\;\;\;\partial_tT^{ti}=-\tau_\text{rel.}^{-1}T^{ti}.
\end{equation}
We can therefore identify $\tau_\text{rel.}$ as the momentum relaxation timescale in the system -- i.e. the characteristic timescale over which the system loses momentum. 

It is a simple task to repeat the previous calculations, with equation (\ref{eq:nonconservationproposal}) replacing the conservation equation, to determine the transverse collective excitations of this effective theory. We find that the dispersion relation of the shear diffusion mode is altered to become
\begin{equation}
\label{eq:modifiedsheardispersionrel}
\omega=-i\left(\tau_\text{rel.}^{-1}+\frac{\eta k^2}{\epsilon+P}\right)+\ldots,
\end{equation}
where the ellipsis represents higher order terms in $k$. In real space, this excitation is a purely decaying mode with decay rate $\Gamma=\tau_\text{rel.}^{-1}+\eta k^2/\left(\epsilon+P\right)$. Our modification of the conservation equation to incorporate the loss of momentum results in the shear diffusion mode decay rate increasing by a constant equal to the inverse of the momentum relaxation timescale (or equivalently, equal to the rate of momentum relaxation). In other words, this excitation radiates its energy faster and thus lives for a shorter time.

We have thus far implicitly assumed that $\tau_\text{rel.}>0$. If in fact $\tau_{\text{rel.}}<0$, the fluid will actually \textit{gain} momentum at a rate proportional to its velocity. Such a state is clearly unstable to any small excitations -- for example, the decay rate of the shear diffusion mode will become negative and thus its amplitude will grow exponentially in time. We therefore demand that $\tau_\text{rel.}\ge0$ for stability.

What we have outlined is, like the usual theory of hydrodynamics, a phenomenological theory -- we have no microscopic description of what is causing the  momentum relaxation of the fluid described here. In a quasiparticle-based description, we can picture it as arising from scattering processes which do not conserve momentum but as usual in holography, a quasiparticle description is probably not an accurate description of the true dynamics of the strongly coupled field theory. 

Furthermore, we emphasise that the modified version of hydrodynamics outlined above is not derived via the usual procedure in hydrodynamics (and in effective field theories more generally) of writing a constitutive relation containing all terms allowed by the symmetries present. While such an analysis would be worthwhile, instead we have just proposed the simplest modification which produces the transverse collective excitation present in the strongly-interacting field theory dual to (\ref{eq:MassiveGravityAction}), which will be derived shortly. This simple modification is the dominant correction to hydrodynamics in this limit. In the field theory dual to (\ref{eq:MassiveGravityAction}), we find that the momentum relaxation rate is given by
\begin{equation}
\tau_\text{rel.}^{-1}=\frac{s\left(m^2_\alpha+m^2_\beta\right)}{2\pi\left(\epsilon+P\right)},
\end{equation}
and that the stability condition $\tau_\text{rel.}\ge0$ is equivalent to the ``wall of stability'' which was previously identified numerically for this theory. There are numerous other ways to test if this modified version of hydrodynamics really is an accurate description of the aforementioned field theory, the most obvious of which is to compute the dispersion relation of the (longitudinal) sound mode in both setups. We leave this to future work.

\subsection{Zero density ``hydrodynamics''}
\label{sec:ZeroDensityhydroAnalytics}

Gauge/gravity duality encodes the collective excitations of the field theory (i.e. the poles of the field theory two-point functions) as the quasinormal modes of the dual bulk gravitational fields. To determine the effect of the diffeomorphism-breaking mass terms $m^2_\alpha,m^2_\beta$ on the shear diffusion mode, we therefore want to calculate the effects of these terms on the corresponding quasinormal mode of the bulk field fluctuations. The simplest case to consider is that of zero density ($\mu=0$), where the fluctuations of the bulk gauge field decouple from those of the bulk metric. In the dual field theory, this means that small fluctuations of energy-momentum decouple from those of the U(1) current and therefore the modified hydrodynamics just introduced -- which assumed that there were no such couplings -- may apply.

In this decoupling limit, it is fluctuations of the transverse components of $T^{ab}$ which support the shear diffusion mode. In this limit, the equations of motion of their dual gravitational fields are
\begin{subequations}
\begin{align}
& \frac{d}{dr}\left[\frac{f}{r^2}\frac{\omega\left(k{h^y_t}'+\omega{h^y_x}'\right)-2f\left(m^2_\beta+m^2_\alpha\frac{r_0}{r}\right){h^y_x}'}{\omega^2-k^2f-2f\left(m^2_\beta+m^2_\alpha\frac{r_0}{r}\right)}\right]+\frac{\omega}{r^2f}\left(\omega h^y_x+kh^y_t\right)-\frac{r_0m_\alpha^2}{r^3}h^y_x=0, \label{eq:ZeroDensitySubstitutedEoMa}\\
& \frac{d}{dr}\left[\frac{f\left(m^2_\beta+m^2_\alpha\frac{r_0}{r}\right)\left(\omega{h^y_t}'+kf{h^y_x}'\right)}{r^2\left[\omega^2-k^2f-2f\left(m^2_\beta+m^2_\alpha\frac{r_0}{r}\right)\right]}\right]+\frac{\omega}{r^2f}\left(m^2_\beta+m^2_\alpha\frac{r_0}{r}\right)h^y_t+\frac{kr_0m^2_\alpha}{2r^3}h^y_x=0.
\label{eq:ZeroDensitySubstitutedEoMb}
\end{align}
\end{subequations}
In this limit, the decoupled equation of motion (\ref{eq:SubstitutedEoM2}) for $a_y$ is invariant under diffeomorphisms (\ref{eq:specificfielddiffeomorphismtransformations}) and will not concern us here. The breaking of diffeomorphism invariance is communicated to it by its coupling to the metric fluctuations at non-zero density.

To compute the dispersion relation of the relevant quasinormal mode, we will divide the spacetime into two overlapping regions -- an ``inner region'' close to the horizon at $r=r_0$ and an ``outer region'' which extends up to the boundary $r=0$. We can then solve the equations of motion in each of these regions up to integration constants. A quasinormal mode must be purely infalling at the horizon: imposing this boundary condition fixes the integration constants of the inner region solutions (up to an overall normalisation of each field). By demanding consistency of the inner and outer region solutions in the matching regions where they overlap, we can then fix the integration constants of the outer region solutions (up to an overall normalisation of each field). To identify a quasinormal mode, we then determine the values of $\omega$ for which these outer region solutions are normalisable at the spacetime boundary.

\subsubsection{Inner region solutions}

The inner region is the region of spacetime close to the horizon defined by $r_0-r\ll r_0$. Expanding the terms in the equations of motion (\ref{eq:ZeroDensitySubstitutedEoMa}) and (\ref{eq:ZeroDensitySubstitutedEoMb}) in this limit, we find that the fields $h^y_t$ and $h^y_x$ decouple, and both obey the equation (assuming that $T\ne0$)
\begin{equation}
\label{eq:ZeroDensityInnerRegion}
\frac{d}{dr}\left[\left(r_0-r\right){h^y_t}'\left(r\right)\right]+\frac{\omega^2}{f'\left(r_0\right)^2\left(r_0-r\right)}h^y_t\left(r\right)=0,
\end{equation}
and similarly for $h^y_x$. To compute the retarded Green's functions of the dual field theory, we need to impose ingoing boundary conditions on these fields at the horizon. After solving (\ref{eq:ZeroDensityInnerRegion}), we impose these ingoing boundary conditions to obtain the following solutions for the fields in the inner region
\begin{equation}
\label{eq:ZeroDensityInnerRegionSolutions}
h^y_x=a_0\exp\left[-\frac{i\omega}{4\pi T}\log\left(\frac{r_0-r}{r_0}\right)\right],\;\;\;\;\;\;\;\;\;\;\;\; h^y_t=b_0\exp\left[-\frac{i\omega}{4\pi T}\log\left(\frac{r_0-r}{r_0}\right)\right],
\end{equation}
where $a_0$ and $b_0$ are integration constants and we have used the expression (\ref{eq:TemperatureDefinition}) for the temperature $T$. This is the standard result for ingoing fields at a horizon of non-zero temperature.

\subsubsection{Outer region solutions}

We define the outer region of the spacetime by the limit
\begin{equation}
r^2\ll\frac{f^2}{\omega^2-k^2f-2f\left(m_\beta^2+m_\alpha^2\frac{r_0}{r}\right)}.
\end{equation}
This region extends to the boundary of the spacetime at $r=0$ but not all the way to the horizon. In this limit, the non-derivative terms in the equations of motion (\ref{eq:ZeroDensitySubstitutedEoMa}) and (\ref{eq:ZeroDensitySubstitutedEoMb}) are subleading and we can neglect them and thus trivially integrate each equation to yield the first order differential equations
\begin{equation}
\begin{aligned}
\label{eq:ZeroDensityOuterRegionSolutions}
&\frac{f}{r^2\left[\omega^2-k^2f-2f\left(m_\beta^2+m_\alpha^2\frac{r_0}{r}\right)\right]}\left[\omega\left(\omega{h^y_x}'+k{h^y_t}'\right)-2f\left(m_\beta^2+m_\alpha^2\frac{r_0}{r}\right){h^y_x}'\right]=\omega kc_1,\\
&\frac{-f\left(m_\beta^2+m_\alpha^2\frac{r_0}{r}\right)}{r^2\left[\omega^2-k^2f-2f\left(m_\beta^2+m_\alpha^2\frac{r_0}{r}\right)\right]}\left(\omega{h^y_t}'+kf{h^y_x}'\right)=\omega c_2,
\end{aligned}
\end{equation}
where $c_1$ and $c_2$ are integration constants that will be fixed shortly. It is simple to decouple the two fields and integrate to give the outer region solutions
\begin{equation}
\begin{aligned}
& h^y_x={h^y_x}^{(0)}+\omega kc_1\int^r_0d\hat{r}\frac{\hat{r}^2}{f\left(\hat{r}\right)}+\omega kc_2\int^r_0d\hat{r}\frac{\hat{r}^2}{f\left(\hat{r}\right)\left[m_\beta^2+m_\alpha^2\frac{r_0}{\hat{r}}\right]},\\
& h^y_t={h^y_t}^{(0)}+\frac{2}{3}c_2r^3-\frac{1}{3}c_1k^2r^3-\omega^2c_2\int^r_0d\hat{r}\frac{\hat{r}^2}{f\left(\hat{r}\right)\left[m_\beta^2+m_\alpha^2\frac{r_0}{\hat{r}}\right]},
\end{aligned}
\end{equation}
where ${h^y_x}^{(0)}$ and ${h^y_t}^{(0)}$ denote the boundary values of the respective fields. These integrals can be done analytically but the results are very lengthy and will not be presented here. Near the boundary of the spacetime, the solutions take the form
\begin{equation}
h^y_x\left(r\right)={h^y_x}^{(0)}+\frac{1}{3}\omega kc_1r^3+\ldots,\;\;\;\;\;\;\;\;\;\;\;\;\;\;\;\;\;\;\;h^y_t\left(r\right)={h^y_t}^{(0)}+\frac{1}{3}\left(2c_2-k^2c_1\right)r^3+\ldots.
\end{equation}
From these expressions, we see that the quasinormal modes are given by the poles of the integration constants $c_1$ and $c_2$. These integration constants can be fixed by demanding that the outer solutions correspond to fields which are ingoing at the horizon, as we will now demonstrate.

\subsubsection{Matching and Green's functions}

Assuming that the momentum and mass terms are of the same order of magnitude as the frequency ($\omega\sim k\sim m_\alpha\sim m_\beta$), the inner and outer regions overlap in a region of spacetime given by
\begin{equation}
\left|\frac{\omega}{f'\left(r_0\right)}\right|\ll\frac{r_0-r}{r_0}\ll1.
\end{equation}
Since $f'\left(r_0\right)\sim T$, this region of spacetime is of a non-zero size provided that $\omega,k,m_\alpha,m_\beta\ll T$. This is the hydrodynamic limit.

To impose ingoing boundary conditions on the outer solutions, we expand both the ingoing inner region solutions (\ref{eq:ZeroDensityInnerRegionSolutions}) and the outer region solutions (\ref{eq:ZeroDensityOuterRegionSolutions}) in this matching region. By demanding that these expansions are consistent, we fix $c_1$ and $c_2$ to be those corresponding to ingoing fields at the horizon. The inner region solutions in the matching region are given by expanding (\ref{eq:ZeroDensityInnerRegionSolutions}) in the limit $\omega\log\left(\frac{r_0-r}{r_0}\right)\ll T$ to give
\begin{equation}
\label{eq:InnerRegionSolutionsMatchingRegionZeroDensity}
h^y_x=a_0\left[1-\frac{i\omega}{4\pi T}\log\left(\frac{r_0-r}{r_0}\right)+\ldots\right],\;\;\;\;\;\;\;\;\;\;\; h^y_t=b_0\left[1-\frac{i\omega}{4\pi T}\log\left(\frac{r_0-r}{r_0}\right)+\ldots\right].
\end{equation}
The outer region solutions in the matching region are found by expanding (\ref{eq:ZeroDensityOuterRegionSolutions}) in the limit $r_0-r\ll r_0$. The constants $c_1$ and $c_2$ are fixed by demanding that the ratios of the coefficients of the $\left(r_0-r\right)^0$ and $\log\left(\frac{r_0-r}{r_0}\right)$ terms in these expansions are the same as that for the inner region solutions in the matching region (\ref{eq:InnerRegionSolutionsMatchingRegionZeroDensity}). This fixes $c_1$ and $c_2$ to be
\begin{equation}
\begin{aligned}
c_1&=\frac{3r_0\left(\omega{h^y_x}^{(0)}+k{h^y_t}^{(0)}\right)+2i\left(m_\alpha^2r_0^2+m_\beta^2r_0^2\right){h^y_x}^{(0)}+\ldots}{kr_0^2\left[k^2r_0^2+2\left(m_\alpha^2r_0^2+m_\beta^2r_0^2\right)-3ir_0\omega+\ldots\right]},\\
c_2&=\frac{-\left(m_\alpha^2+m_\beta^2\right)\left(3{h^y_t}^{(0)}-ikr_0{h^y_x}^{(0)}\right)+\ldots}{r_0\left[k^2r_0^2+2\left(m^2_\alpha r_0^2+m_\beta^2r_0^2\right)-3ir_0\omega+\ldots\right]},
\end{aligned}
\end{equation}
where the ellipses denote higher order terms in $\omega,k,m_\alpha,m_\beta$. 

These constants exhibit a pole at a given $\omega\left(k,m_\alpha,m_\beta,T\right)$ which is the dispersion relation of the quasinormal mode of the dual gravitational solution as explained in the previous subsection. This coincides with the dispersion relation of the poles of the field theory retarded Green's functions. To compute these Green's functions, we evaluate the on-shell action (\ref{eq:onshellactionfullformula}) to give (at lowest order in $\omega,k,m_\alpha,m_\beta$)
\begin{equation}
\begin{aligned}
S=\frac{L^2}{4\kappa_4^2r_0^3}\int&\frac{d\omega dk}{\left(2\pi\right)^2}\frac{1}{k^2r_0^2+2\left(m_\alpha^2 r_0^2+m_\beta^2r_0^2\right)-3ir_0\omega}\Bigl\{\left[-3r_0^2k^2-6\left(m_\alpha^2 r_0^2+m_\beta^2r_0^2\right)\right]{h^y_t}^{(0)}{h^y_t}^{(0)}\\
&-3r_0^2\omega k\left[{h^y_x}^{(0)}{h^y_t}^{(0)}+{h^y_t}^{(0)}{h^y_x}^{(0)}\right]+\left[-3r_0^2\omega^2-2ir_0\omega\left(m_\alpha^2r_0^2+m_\beta^2r_0^2\right)\right]{h^y_x}^{(0)}{h^y_x}^{(0)}\Bigr\},
\end{aligned}
\end{equation}
where ${h^y_t}^{(0)}{h^y_t}^{(0)}$ is shorthand for ${h^y_t}^{(0)}\left(-\omega,-k\right){h^y_t}^{(0)}\left(\omega,k\right)$ and similarly for the other terms with superscript $(0)$, and where we have neglected terms which are analytic in $\omega,k$ and thus produce contact terms in the dual field theory Green's functions.

Using the usual holographic prescription \cite{Son:2002sd,Kaminski:2009dh}, we find that the retarded Green's functions of the dual field theory in the limit $\omega,k,m_\alpha,m_\beta\ll T$ are (up to contact terms)
\begin{equation}
\begin{aligned}
G^R_{T^{ty}T^{ty}}\left(\omega,k\right)&=\frac{L^2}{2\kappa_4^2r_0^3}\frac{3r_0^2k^2+6\left(m_\alpha^2r_0^2+m_\beta^2r_0^2\right)}{3ir_0\omega-k^2r_0^2-2\left(m_\alpha^2r_0^2+m_\beta^2r_0^2\right)},\\
G^R_{T^{xy}T^{xy}}\left(\omega,k\right)&=\frac{L^2}{2\kappa_4^2r_0^3}\frac{3r_0^2\omega^2+2ir_0\omega\left(m_\alpha^2r_0^2+m_\beta^2r_0^2\right)}{3ir_0\omega-k^2r_0^2-2\left(m_\alpha^2r_0^2+m_\beta^2r_0^2\right)},\\
G^R_{T^{ty}T^{xy}}\left(\omega,k\right)&=G^R_{T^{xy}T^{ty}}\left(\omega,k\right)=\frac{L^2}{2\kappa_4^2r_0^3}\frac{3r_0^2\omega k}{3ir_0\omega-k^2r_0^2-2\left(m_\alpha^2r_0^2+m_\beta^2r_0^2\right)}.
\end{aligned}
\end{equation}
These Green's functions have two important features. Firstly, they no longer satisfy the Ward identities due to momentum conservation (\ref{eq:wardidentities}), as we anticipated. Secondly, the Green's functions have a pole with dispersion relation
\begin{equation}
\begin{aligned}
\label{eq:ZeroDensityDispersionRelation}
\omega&=-i\frac{r_0k^2}{3}-i\frac{2r_0}{3}\left(m_\alpha^2+m_\beta^2\right)+\ldots\\
&=-i\left(\frac{\eta k^2}{\epsilon+P}+\tau_\text{rel.}^{-1}\right)+\ldots,
\end{aligned}
\end{equation}
where we have defined
\begin{equation}
\label{eq:ourvalueoftau}
\tau_\text{rel.}^{-1}=\frac{s\left(m_\alpha^2+m_\beta^2\right)}{2\pi\left(\epsilon+P\right)},
\end{equation}
and where we have used the expressions
\begin{equation}
\epsilon=\frac{L^2}{\kappa_4^2r_0^3},\;\;\;\;\;\;\;\;\;\;\;\; P=\frac{\epsilon}{2},\;\;\;\;\;\;\;\;\;\;\;\; \eta=\frac{s}{4\pi}=\frac{L^2}{2\kappa_4^2r_0^2},
\end{equation}
for the relevant thermodynamic and transport quantities when $\mu=m_\alpha=m_\beta=0$. Although $m_\alpha$ and $m_\beta$ are non-vanishing in our calculation, they are small with respect to $T$ and hence their effect on these thermodynamic quantities produces contributions to the dispersion relation which are subleading. Note that we have rewritten the $2r_0/3$ factor in $\tau_\text{rel.}^{-1}$ in terms of thermodynamic quantities. Written in this way, we will shortly show that this relation holds also at non-zero densities.

This dispersion relation is precisely that expected from the modified version of hydrodynamics we described in section \ref{sec:ModifiedHydrodynamicsSection} and we therefore interpret $\tau^{-1}_\text{rel.}$ as the timescale of momentum relaxation in the field theory. Looking back on the bulk equations of motion (\ref{eq:ZeroDensitySubstitutedEoMa}) and (\ref{eq:ZeroDensitySubstitutedEoMb}), one can see from the denominators of each equation that the effective momentum of each field near the horizon (where dissipation occurs) is $k^2_\text{eff.}=k^2+2\left(m_\alpha^2+m_\beta^2\right)$ and so the contributions of $m_\alpha^2$ and $m_\beta^2$ to the dispersion relation (\ref{eq:ZeroDensityDispersionRelation}) are perhaps not too surprising. 

Finally, we note that the mode (\ref{eq:ZeroDensityDispersionRelation}) crosses to the upper half-plane and therefore is unstable when $\tau_\text{rel.}^{-1}<0$ (and $k=0$). In terms of the parameters of the original action, this condition for stability may be written
\begin{equation}
\frac{F^2m^2s}{\left(\epsilon+P\right)}\left(\beta+\frac{\alpha L}{2r_0F}\right)\le0,
\end{equation}
which coincides precisely with the ``wall of stability'' identified numerically in \cite{Vegh:2013sk}. As indicated previously, this is a homogeneous (i.e. $k=0$) instability which occurs because the system begins to gain (rather than lose) momentum at a constant rate such that the momentum of the system grows exponentially in time.

\subsection{Non-zero density ``hydrodynamics'' and the Drude model}

Let us now generalise our previous discussion of modified hydrodynamics to a theory with a conserved U(1) charge. When such a charge is present, we must supplement the previous hydrodynamic equations by a constitutive relation for the U(1) current $J^a$ and the conservation equation $\partial_aJ^a=0$. We will consider a current with the usual constitutive relation of that of a conformal theory
\begin{equation}
\label{eq:constitutiverelationconformalcurrent}
J^a=\rho u^a+\ldots,
\end{equation}
where $\rho$ is the charge density and the ellipsis denotes terms of higher order in spacetime derivatives. To compute the collective excitations, we can consider the near-equilibrium configuration (\ref{eq:neareqmfluidflow}) with $\rho\rightarrow\rho+\delta\rho$ also. The equations of motion for $J^a$ and $T^{ab}$ are now coupled but for transverse fluid flow (only $\delta u_y\ne0$), the conservation equation for $J^a$ is identically zero and thus the collective mode dispersion relation is (\ref{eq:modifiedsheardispersionrel}), the same as before. 

There is one difference with respect to the zero charge density case which is that the collective shear diffusion mode, which is an eigenmode with $\delta u_y\ne0$, now results in non-trivial flow of current $J^y$ in the transverse direction (as can be seen from equation (\ref{eq:constitutiverelationconformalcurrent})) in addition to a non-trivial flow of transverse energy-momentum. In the hydrodynamic limit of a field theoretical system, this results in the two-point functions of $J^y,T^{ty}$ and $T^{xy}$ all having a pole with the dispersion relation of this collective mode.

This case of a system with a non-vanishing conserved U(1) charge is relevant for the strongly coupled field theory dual to (\ref{eq:MassiveGravityAction}) when $\mu\ne0$.\footnote{Note that although we have broken bulk diffeomorphism invariance, we have not broken bulk U(1) gauge invariance and so the U(1) current in the dual field theory is still conserved.} Our modified version of hydrodynamics predicts a shear diffusion mode with dispersion relation
\begin{equation}
\begin{aligned}
\label{eq:nonzerodensitydispersionrelation}
\omega&=-i\frac{\eta}{\epsilon+P}k^2-i\tau_\text{rel.}^{-1}+\ldots,\\
&=-i\frac{r_0}{3\left(1+\frac{1}{4}\mu^2r_0^2\right)}\left[k^2+2\left(m^2_\alpha+m^2_\beta\right)\right]+\ldots,
\end{aligned}
\end{equation}
where we have used the result (\ref{eq:ourvalueoftau}) for $\tau_\text{rel.}^{-1}$ and the following thermodynamic quantities valid in the $m^2_\alpha=m^2_\beta=0$ limit
\begin{equation}
\epsilon=\frac{L^2}{\kappa_4^2r_0^3}\left(1+\frac{1}{4}\mu^2 r_0^2\right),\;\;\;\;\;\;\;\;\;\;\;\; P=\frac{\epsilon}{2},\;\;\;\;\;\;\;\;\;\;\;\; \eta=\frac{s}{4\pi}=\frac{L^2}{2\kappa_4^2r_0^2}.
\end{equation}
As before, the corrections to these due to non-zero $m^2_\alpha,m^2_\beta$ will produce subleading terms in the dispersion relations. 

\begin{figure*}
\begin{center}
\includegraphics[scale=1.00]{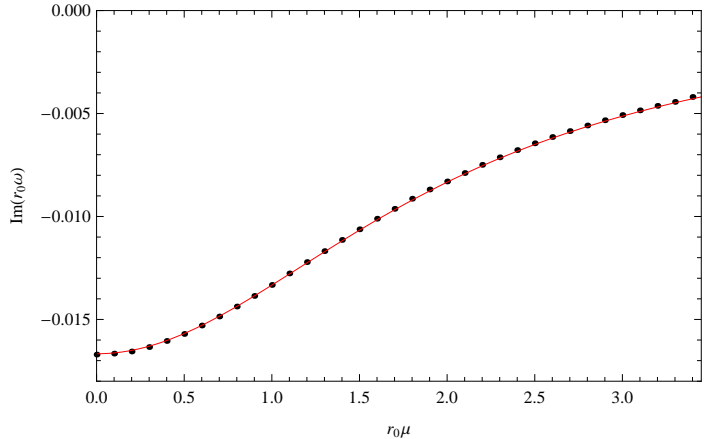}
\caption{Dependence of the imaginary part of the shear diffusion mode upon the field theory chemical potential $\mu$, for $r_0^2m_\alpha^2=r_0^2m_\beta^2=0.01$. The real part (not shown) is zero. The black dots are the results from numerically integrating the equations of motion, and the solid red line is the expression (\ref{eq:nonzerodensitydispersionrelation}).}
\label{fig:NonZeroDensityPoleLocation}
\end{center}
\end{figure*}

To check the validity of this result at non-zero densities, we wish to compute the poles of the retarded Green's functions of $J^y, T^{ty}$ and $T^{xy}$ by determining the quasinormal modes of the dual bulk fields. The extra coupling between $a_y$ and the metric fluctuations $h^y_t,h^y_x$ when $\mu\ne0$ makes this a complicated task. We therefore computed the poles of these Green's functions numerically by integrating the equations of motion (\ref{eq:SubstitutedEoM1}), (\ref{eq:SubstitutedEoM2}) and (\ref{eq:SubstitutedEoM3}) from the horizon (where we applied ingoing boundary conditions) to the boundary for various $\omega$ and searching for the existence of a quasinormal mode using the procedure described in \cite{Kaminski:2009dh}. Note that the Green's functions all share a common set of poles as their dual bulk fields are coupled. Working in the limit $\omega,k,m_\alpha,m_\beta\ll T$, we found a quasinormal mode with dispersion relation (\ref{eq:nonzerodensitydispersionrelation}) for all values of $\mu$ that we studied. A sample of these results are shown in figure \ref{fig:NonZeroDensityPoleLocation}.

In summary, we have therefore shown that our modified version of hydrodynamics -- in which we accounted for the relaxation of momentum over a single characteristic timescale $\tau_\text{rel.}$-- is an accurate description of the field theory dual to the massive gravity theory (\ref{eq:MassiveGravityAction}) with $\mu\ne0$ in the hydrodynamic limit $\omega,k,m_\alpha,m_\beta\ll T$. As in the $\mu=0$ case, there is an instability when $\tau_\text{rel.}<0$ due to momentum absorption. This coincides with the ``wall of stability'' found numerically in \cite{Vegh:2013sk}.

In fact, the charge transport in this modified hydrodynamic theory is essentially the same as that of the simple Drude model. At low enough energies and momenta we can approximate the full two-point function of $J^y$ by just the contribution from the shear diffusion pole. Using the Kubo formula, we can then determine the conductivity
\begin{equation}
\sigma\left(\omega\right)\equiv\frac{i}{\omega}G^R_{J^yJ^y}\left(\omega,k=0\right)\sim\frac{1}{\tau_\text{rel.}^{-1}-i\omega}\sim\frac{1}{1-i\omega\tau_{\text{rel.}}},
\end{equation}
where we do not have an analytic expression for the numerator and we have not included corrections that are higher order in $\omega$. This has the same form as the conductivity computed using the Drude model in which the current is composed of charged, classical particles which, in the absence of external forces, undergo collisions with ions with a mean free time $\tau_{\text{rel.}}$ (see \cite{Vegh:2013sk} for a brief review of this). This result clearly applies more generally than in the specific microscopic model of Drude, as we have shown here, and relies only on the hydrodynamic structure just outlined.

\section{The zero temperature conductivity: beyond the Drude model}
\label{sec:ZeroTemperatureConductivityAnalysis}

As mentioned in the introduction, it is of interest to study non-Drude behaviour of the conductivity in light of unexplained experimental measurements of the conductivity in the normal phase of certain high-$T_c$ superconductors \cite{experimentalpaper1}. These interesting measurements were performed at low temperatures $T<\omega$ outside the hydrodynamic limit, and furthermore it was shown in \cite{Vegh:2013sk} that the field theory dual to (\ref{eq:MassiveGravityAction}) exhibits such non-Drude behaviour at low temperatures. In this section we will compute the conductivity of this theory when $T=0$ and determine analytically the corrections to the simple Drude model. We will first briefly recall the important features of the conductivity for the massless case where momentum is conserved.

The conductivity can be determined from the retarded Green's function of the transverse current via the Kubo formula
\begin{equation}
\label{eq:kuboformulaconductivity}
\sigma\left(\omega\right)\equiv\frac{i}{\omega}G^R_{J^yJ^y}\left(\omega,k=0\right).
\end{equation}
When $m^2_\alpha=m^2_\beta=0$, the leading behaviour of the conductivity at small $\omega$ (neglecting the delta function which is present due to translational invariance) is $\text{Re}\left(\sigma\right)\propto\omega^2,\;\text{Im}\left(\sigma\right)\propto\omega^{-1}$. The scaling of $\text{Im}\left(\sigma\right)$ simply indicates that there must be a delta function in $\text{Re}\left(\sigma\right)$, due to the Kramers-Kronig relation \cite{Hartnoll:2009sz}. This delta function has a simple physical interpretation: because momentum is conserved, the charges will accelerate indefinitely under an applied field and thus the DC conductivity is infinite. Neglecting this delta function, the scaling of $\text{Re}\left(\sigma\right)$ is controlled by the dimension of a scalar operator in the 1-dimensional CFT that governs the theory at low energies as we will now briefly review.

In the massless case \cite{Edalati:2009bi}, the bulk field $a_y$ dual to $J^y$  decouples from the bulk metric fluctuations when $k=0$ and can therefore be treated independently. Near the horizon, this field behaves like a scalar field in AdS$_2$ with mass $(mL_2)^2=2$ where $L_2=L/\sqrt{6}$ is the radius of the near-horizon AdS$_2$ geometry. Utilising the AdS/CFT correspondence, we then expect that the low-energy dissipative properties of the two-point function of $J^y$ are controlled by the two-point function of a scalar operator of dimension $\Delta=2$ in the CFT$_1$ dual to the near-horizon AdS$_2$ geometry. This expectation is borne out by a careful calculation of the conductivity, the result of which is \cite{Edalati:2009bi}
\begin{equation}
\sigma\left(\omega\right)\propto\frac{i}{\omega}\frac{1+t_0\mathcal{G}_2\left(\omega\right)+\ldots}{1+t_1\mathcal{G}_2\left(\omega\right)+\ldots}\;\;\;\;\;\;\;\;\longrightarrow\;\;\;\;\;\;\;\text{Re}\left(\sigma\right)\propto\omega^{2}+\ldots,\;\;\;\;\;\text{Im}\left(\sigma\right)\propto\omega^{-1}+\ldots,
\end{equation}
where $\mathcal{G}_{2}\left(\omega\right)\sim i\omega^3$ is the Greens function of a scalar operator of dimension $2$ in the CFT$_1$ dual to the near-horizon AdS$_2$ region, $t_i$ are $\omega$-independent constants and ellipses denote higher order terms in a small $\omega$ expansion. As previously stated, the power of the leading term in $\text{Re}\left(\sigma\right)$ at small $\omega$ is controlled by the CFT$_1$ dual to the near-horizon geometry. A similar power law holds for many other holographic theories \cite{Goldstein:2009cv,Hartnoll:2010gu}.

When $m_\alpha^2,m_\beta^2\ne0$, our background solution (\ref{eq:backgroundsolution}) still possesses an AdS$_2$ near-horizon geometry. However, the bulk fields $a_y$ and $h^y_t$ are now coupled. This is a consequence of the extra dynamical bulk degrees of freedom present when the graviton has a mass. The result is that, at low energies, the current $J^y$ couples to two scalar operators in the CFT$_1$ -- one with $\Delta=2$ (as in the massless case), and one with $\Delta=1$. In the limit $\omega\rightarrow0$, the coupling to the scalar operator with $\Delta=1$ will dominate and thus $\text{Re}\left(\sigma\right)\propto\omega^{-1}\mathcal{G}_1\left(\omega\right)=\omega^0$, where $\mathcal{G}_1\left(\omega\right)\sim i\omega$ is the two-point function of a scalar operator with $\Delta=1$ in the CFT$_1$. This results in a finite DC conductivity at zero temperature.

\subsection{The zero momentum master fields}

To compute the conductivity analytically, we will restrict ourselves to the case $m_\alpha=0$. In this limit, we can write the $k=0$ equations of motion involving $a_y$ as two decoupled equations for two ``master fields'', for all values of $r$. These decoupled ``master fields'' are given by
\begin{equation}
\label{eq:nomomentummasterfields}
\varphi_{\pm}\left(r\right)=a_y+\gamma_{\pm}\left[rA_t'a_y+\frac{1}{r}\left({h^y_t}'+i\omega h^y_r\right)\right],
\end{equation}
where
\begin{equation}
\gamma_{\pm}=-\frac{3}{4m_\beta^2\mu r_0^2}\left[\left(1-r_0^2m_\beta^2+\frac{1}{4}r_0^2\mu^2\right)\pm\sqrt{\left(1-r_0^2m_\beta^2+\frac{1}{4}r_0^2\mu^2\right)^2+\frac{8}{9}m_\beta^2\mu^2r_0^4}\right],
\end{equation}
and they obey the equations of motion
\begin{equation}
\label{eq:nomomentumequations}
\begin{aligned}
\frac{d}{dr}\left[f(r)\varphi_{\pm}'\left(r\right)\right]+\varphi_{\pm}\left(r\right)\Biggl\{\frac{\omega^2}{f}-\frac{r^2\mu^2}{r_0^2}-\frac{2m_\beta^2\mu\gamma_{\pm}r}{r_0}\Biggr\}=0.
\end{aligned}
\end{equation}
These equations are derived in the appendix. In the $T=0$ limit, 
\begin{equation}
\label{eq:ZeroTemperatureEquationforMu}
r_0\mu=2\sqrt{3-\left(r_0m_\beta\right)^2},
\end{equation}
and thus the equations of motion simplify to
\begin{equation}
\label{eq:zerotemperaturemasterequations}
\frac{d}{dr}\left[f\varphi_\pm'(r)\right]+\varphi_\pm\left(r\right)\left\{\frac{\omega^2}{f}+\frac{3r}{r_0^3}\left[2-r_0^2m_\beta^2\pm\left(2-\frac{1}{3}r_0^2m_\beta^2\right)\right]-\frac{4\left(3-r_0^2m_\beta^2\right)r^2}{r_0^4}\right\}=0,
\end{equation}
for the fields
\begin{equation}
\label{eq:zerotemperaturemastervariables}
\varphi_{\pm}=a_y-\frac{3\left[2-r_0^2m_\beta^2\pm\left(2-\frac{1}{3}r_0^2m^2_\beta\right)\right]}{4r_0m_\beta^2\sqrt{3-r_0^2m_\beta^2}}\left\{rA_t'a_y+\frac{1}{r}\left({h^y_t}'+i\omega h^y_r\right)\right\}.
\end{equation}

At $T=0$ our background solution has one dimensionless parameter, $m_\beta^2r_0^2$, which is the ratio of the graviton mass to the field theory chemical potential, or equivalently $\mu\tau_\text{rel.}$ in the dual field theory as we will see shortly. We restrict to $m_\beta^2r_0^2\leq3$ so that $\mu$, given in (\ref{eq:ZeroTemperatureEquationforMu}), is always real. 

To compute the conductivity, we will use a similar strategy to that used for the massless theory \cite{Edalati:2009bi}. We divide the spacetime into inner and outer regions, solve the equations in each region, and match them in an overlapping region to fix the integration constants of the outer region solutions such that the fields are ingoing at the horizon. We can then read off the values of the fields at the boundary and, from the on-shell action, compute the Green's functions and hence the conductivity from the Kubo formula (\ref{eq:kuboformulaconductivity}).

\subsection{Solutions in the inner AdS$_2$ region and beyond}

When $T=0$, the geometry (\ref{eq:backgroundsolution}) has a near-horizon AdS$_2$ region which we expect will control the low energy, dissipative properties of the dual field theory. To see this near-horizon region at the level of the equations of motion, we change the radial co-ordinate to \cite{Faulkner:2009wj,Edalati:2009bi}
\begin{equation}
\label{eq:ZetaCoordinateDefinition}
\zeta=\frac{\omega r_0}{\left(6-m_\beta^2r_0^2\right)}\frac{r_0}{\left(r_0-r\right)}.
\end{equation}
After writing the equations of motion (\ref{eq:zerotemperaturemasterequations}) in this new radial co-ordinate, we can expand at small $\omega$ (keeping $\zeta$ fixed) to give
\begin{equation}
\begin{aligned}
\label{eq:zerotemperatureinnerequations}
&\varphi_+''\left(\zeta\right)+\varphi_+'\left(\zeta\right)\left[\frac{2\left(4-r_0^2m_\beta^2\right)r_0\omega}{\left(6-r_0^2m_\beta^2\right)^2\zeta^2}\right]+\varphi_+\left(\zeta\right)\left[1+\frac{4\left[3-r_0^2m_\beta^2+\zeta^2\left(4-r_0^2m_\beta^2\right)\right]r_0\omega}{\left(6-r_0^2m_\beta^2\right)^2\zeta^3}\right]=0,\\
&\varphi_-''\left(\zeta\right)+\varphi_-'\left(\zeta\right)\left[\frac{2\left(4-r_0^2m_\beta^2\right)r_0\omega}{\left(6-r_0^2m_\beta^2\right)^2\zeta^2}\right]+\varphi_-\left(\zeta\right)\left[1-\frac{2}{\zeta^2}+\frac{2\left(4-r_0^2m_\beta^2\right)\left(1+2\zeta^2\right)r_0\omega}{\left(6-r_0^2m_\beta^2\right)^2\zeta^3}\right]=0,
\end{aligned}
\end{equation}
where we have neglected terms of order $\omega^2$ and higher. This expansion is valid very close to the horizon of the spacetime with $\zeta$, which is roughly $r_0^2\omega/\left(r_0-r\right)$, kept fixed. At very low energies in this inner region, we can set $\omega=0$ in (\ref{eq:zerotemperatureinnerequations}) and find that (after rescaling $\zeta$ by a factor of $\omega$) $\varphi_+$ and $\varphi_-$ obey the equations of motion of scalar fields of mass $\left(mL_2\right)^2=0,2$ in an AdS$_2$ spacetime with radius $L_2=L/\sqrt{6-m_\beta^2r_0^2}$. Therefore, with a non-zero mass $m_\beta$, there is still an emergent CFT$_1$ controlling the low-energy dissipative dynamics of the field theory. 

However, the mass term $m_\beta$ does have one important consequences for these dynamics. The dissipative dynamics of $J^y$ are now controlled by \textit{two} operators in this CFT$_1$ -- one of dimension $\Delta=1$ and one of dimension $\Delta=2$ -- since both of $\varphi_\pm$ depend upon $a_y$. It is the coupling to this dimension 1 operator that will produce a finite DC conductivity. 

We can solve the inner region equations (\ref{eq:zerotemperatureinnerequations}) perturbatively in $\omega$. After imposing ingoing boundary conditions on the solutions at the horizon  $\zeta\rightarrow\infty$, these are
\begin{equation}
\begin{aligned}
\label{eq:zerotempinnerregionsolns}
\varphi_+\left(\zeta\right)&=e^{i\zeta}\left\{a_0+a_1\omega+\frac{2a_0r_0\omega}{\zeta\left(6-m_\beta^2r_0^2\right)^2}\left[-\left(3-m_\beta^2r_0^2\right)+i\left(4-m_\beta^2r_0^2\right)\zeta\log\zeta\right]\right\}\\
&+e^{-i\zeta}\frac{4ia_0\omega r_0\left(3-r_0^2m_\beta^2\right)}{\left(6-m_\beta^2r_0^2\right)^2}\left[\text{Ei}\left(2i\zeta\right)-i\pi\right]+O\left(\omega^2\right),\\
\varphi_-\left(\zeta\right)&=b_0\left(1+\frac{i}{\zeta}\right)e^{i\zeta}+O\left(\omega\right),
\end{aligned}
\end{equation}
where $\text{Ei}\left(z\right)$ is the exponential integral \cite{AbramowitzStegun} and $a_0$, $a_1$ and $b_0$ are integration constants. Because $\varphi_-$ is dual to an operator of higher dimension than $\varphi_+$, its contribution to the conductivity at low $\omega$ will be subleading, and therefore we only require it to leading order in $\omega$ as shown above.

Note that the solutions (\ref{eq:zerotempinnerregionsolns}), which include corrections of order $\omega^1$, go beyond the simple form of scalar fields in AdS$_2$ which are valid to order $\omega^0$. These solutions are sensitive not just to the AdS$_2$ geometry infinitesimally close to the horizon but also to corrections to this geometry encapsulated by the order $\omega^1$ corrections.

\subsection{Solutions in the outer region}

We can define an outer region of the spacetime, which includes the boundary at $r=0$, by the inequality $r^2\omega^2\ll f^2$. In this limit, the equations of motion (\ref{eq:zerotemperaturemasterequations}) become
\begin{equation}
\frac{d}{dr}\left[f\varphi_\pm'(r)\right]+\varphi_\pm\left(r\right)\left\{\frac{3r}{r_0^3}\left[2-r_0^2m_\beta^2\pm\left(2-\frac{1}{3}r_0^2m_\beta^2\right)\right]-\frac{4\left(3-r_0^2m_\beta^2\right)r^2}{r_0^4}\right\}=0,
\end{equation}
and can be solved analytically to give
\begin{equation}
\begin{aligned}
\label{eq:zerotempouterregionsolutionplus}
\varphi_+=&\left[1+\frac{2\left(3-m_\beta^2r_0^2\right)r}{m_\beta^2r_0^3}\right]\Biggl\{\varphi_+^{(0)}+\theta_+\Biggl[\frac{r_0\left(6-m_\beta^2r_0^2\right)}{r_0-r}-\frac{8r_0\left(6-m_\beta^2r_0^2\right)\left(3-m_\beta^2r_0^2\right)^2}{\left(2-m_\beta^2r_0^2\right)\left[2\left(3-m_\beta^2r_0^2\right)r+m_\beta^2r_0^3\right]}\\
&-\left(6-m_\beta^2r_0^2\right)+\frac{8\left(6-m_\beta^2r_0^2\right)\left(3-m_\beta^2r_0^2\right)^2}{m_\beta^2r_0^2\left(2-m_\beta^2r_0^2\right)}-2\left(10-3m_\beta^2r_0^2\right)\log\left(\frac{r_0-r}{r_0}\right)\\
&+\frac{-m_\beta^6r_0^6-3m_\beta^4r_0^4+48m_\beta^2r_0^2-92}{\left(2-m_\beta^2r_0^2\right)^\frac{3}{2}}\Biggl\{\tan^{-1}\left(\frac{\left(3-m_\beta^2r_0^2\right)r+r_0}{r_0\sqrt{2-m_\beta^2r_0^2}}\right)-\tan^{-1}\left(\frac{1}{\sqrt{2-m_\beta^2r_0^2}}\right)\Biggr\}\\
&+\left(10-3m_\beta^2r_0^2\right)\log\left[\frac{\left(3-m_\beta^2r_0^2\right)r^2+2rr_0+r_0^2}{r_0^2}\right]\Biggr]\Biggr\},
\end{aligned}
\end{equation}
and
\begin{equation}
\begin{aligned}
\label{eq:zerotempouterregionsolutionminus}
\varphi_-=&\frac{r_0-r}{r_0}\Biggl\{\varphi_-^{(0)}+\theta_-\Biggl[r_0\left(6-m_\beta^2r_0^2\right)\left(13m_\beta^4r_0^4-111m_\beta^2r_0^2+246\right)\\
&+12r_0\left(-m_\beta^6r_0^6+11m_\beta^4r_0^4-42m_\beta^2r_0^2+56\right)\log\left(\frac{r_0-r}{r_0}\right)\\
&-6r_0\left(-m_\beta^6r_0^6+11m_\beta^4r_0^4-42m_\beta^2r_0^2+56\right)\log\left[\frac{\left(3-m_\beta^2r_0^2\right)r^2+2rr_0+r_0^2}{r_0^2}\right]\\
&-\frac{3r_0\left(m_\beta^8r_0^8-10m_\beta^6r_0^6+37m_\beta^4r_0^4-68m_\beta^2r_0^2+68\right)}{\sqrt{2-m_\beta^2r_0^2}}\Biggl[\tan^{-1}\left(\frac{\left(3-m_\beta^2r_0^2\right)r+r_0}{r_0\sqrt{2-m_\beta^2r_0^2}}\right)\\
&-\tan^{-1}\left(\frac{1}{\sqrt{2-m_\beta^2r_0^2}}\right)\Biggr]\Biggr]\Biggr\}\\
&+\frac{1}{\left(r_0-r\right)^2}\Biggl\{-\left(6-m_\beta^2r_0^2\right)r_0\theta_-\Bigl[3r^2\left(3m_\beta^4r_0^4-23m_\beta^2r_0^2+46\right)\\
&-3rr_0\left(7m_\beta^4r_0^4-56m_\beta^2r_0^2+116\right)+r_0^2\left(13m_\beta^4r_0^4-111m_\beta^2r_0^2+246\right)\Bigr]\Biggr\},
\end{aligned}
\end{equation}
where $\varphi_\pm^{(0)}$ are the values of $\varphi_\pm$ at $r=0$ and $\theta_\pm$ are integration constants.

\subsection{Matching and the conductivity}

To fix the integration constants $\theta_\pm$ such that the fields are ingoing at the horizon, we expand the outer region solutions (\ref{eq:zerotempouterregionsolutionplus}) and (\ref{eq:zerotempouterregionsolutionminus}) near the horizon $r_0-r\ll r_0$, and the ingoing inner region solutions (\ref{eq:zerotempinnerregionsolns}) far from the horizon $\zeta\rightarrow0$. In this `matching region' given by
\begin{equation}
\label{eq:zeroTmatchingregionconditions}
\frac{\omega r_0}{6-m_\beta^2r_0^2}\ll\frac{r_0-r}{r_0}\ll1,
\end{equation}
we can fix $\theta_\pm$ by matching the terms of order $\left(r-r_0\right)^{-1}$, $\log\left(\frac{r_0-r}{r_0}\right)$ and $\left(r_0-r\right)^0$ in $\varphi_+$ and the terms of order $\left(r_0-r\right)^{-2}$ and $\left(r_0-r\right)^1$ in $\varphi_-$. This yields
\begin{equation}
\begin{aligned}
\theta_+\left(\omega\right)=&\frac{i\omega r_0\varphi_+^{(0)}+\ldots}{\left(6-m_\beta^2r_0^2\right)^2}\left[1+\kappa_1i\omega\log\left(\frac{\omega r_0}{6-m_\beta^2r_0^2}\right)+\kappa_2\omega+\kappa_3i\omega+\ldots\right]^{-1},\\
\theta_-\left(\omega\right)=&-\frac{i\omega^3r_0^2\varphi_-^{(0)}}{3\left(6-m_\beta^2r_0^2\right)^6}\left[1+\ldots\right],
\end{aligned}
\end{equation}
where the ellipses denote higher order terms in $\omega$ and the constants $\kappa_i$ in $\theta_+\left(\omega\right)$ are
\begin{equation}
\begin{aligned}
\label{eq:kappadefinitions}
\kappa_1=&\frac{8r_0\left(3-m_\beta^2r_0^2\right)}{\left(6-m_\beta^2r_0^2\right)^2},\;\;\;\;\;\;\;\;\;\;\;\;\;\;\;\;\kappa_2=\frac{4\pi r_0\left(3-m_\beta^2r_0^2\right)}{\left(6-m_\beta^2r_0^2\right)^2},\\
\kappa_3=&\frac{r_0}{\left(6-m_\beta^2r_0^2\right)^2}\Biggl[\left(3-m_\beta^2r_0^2\right)\left(8\gamma_E+8\log2-9\right)-\frac{16\left(3-m_\beta^2r_0^2\right)^3}{m_\beta^2r_0^2\left(2-m_\beta^2r_0^2\right)}+3\left(4-m_\beta^2r_0^2\right)\\
&-\frac{-m_\beta^6r_0^6-3m_\beta^4r_0^4+48m_\beta^2r_0^2-92}{\left(2-m_\beta^2r_0^2\right)^\frac{3}{2}}\Biggl\{\tan^{-1}\left(\frac{4-m_\beta^2r_0^2}{\sqrt{2-m_\beta^2r_0^2}}\right)-\tan^{-1}\left(\frac{1}{\sqrt{2-m_\beta^2r_0^2}}\right)\Biggr\}\\
&-\left(10-3m_\beta^2r_0^2\right)\log\left(6-m_\beta^2r_0^2\right)\Biggr],
\end{aligned}
\end{equation}
with $\gamma_E$ the Euler-Mascheroni constant. $\theta_\pm\left(\omega\right)$ are expansions in the quantity $\omega r_0/\left(6-r_0^2m_\beta^2\right)^2$, which, from (\ref{eq:zeroTmatchingregionconditions}), must be small for the matching to be accurate.

To determine the two-point function of $J^y$, we need to know the value of the on-shell action for a bulk field configuration satisfying $a_y^{(0)}=1, {h^y_t}^{(0)}=0$ where the superscript $(0)$ denotes the boundary value of the corresponding field. By comparing the expansions of the solutions for $\varphi_\pm$ (given in (\ref{eq:zerotempouterregionsolutionplus}) and (\ref{eq:zerotempouterregionsolutionminus})) around $r=0$ with the expression obtained by expanding the fundamental fields $a_y$ and $h^y_t$ around $r=0$ and substituting into the definition (\ref{eq:zerotemperaturemastervariables}) of $\varphi_\pm$, we can solve for the coefficients in the expansions of $a_y$ and $h^y_t$ in terms of the coefficients in the expansions of $\varphi_\pm$. After substituting the expansions of these fundamental fields into the on-shell action (\ref{eq:onshellactionfullformula}), we find that the on-shell action for a field configuration with ${h^y_t}^{(0)}=0$ is
\begin{equation}
\label{eq:zeroTonshellactionresult}
S=\frac{L^2}{4\kappa_4^2}\int\frac{d\omega}{2\pi}\frac{\left(6-m_\beta^2r_0^2\right)^3}{r_0}\left[-\frac{1}{m_\beta^2r_0^2}\frac{\theta_+\left(\omega\right)}{\varphi_+^{(0)}}+6r_0\left(3-m_\beta^2r_0^2\right)\frac{\theta_-\left(\omega\right)}{\varphi_-^{(0)}}\right]a_y^{(0)}\left(-\omega\right)a_y^{(0)}\left(\omega\right).
\end{equation}
At this point we will ignore the contribution from $\theta_-\left(\omega\right)$ -- it is suppressed by a power of $\omega^2$ with respect to $\theta_+\left(\omega\right)$ in the small $\omega$ limit where our calculation is valid. Since we have neglected terms of this order in the expansion of $\theta_+\left(\omega\right)$, we should neglect the contribution of $\theta_-\left(\omega\right)$ for consistency. The leading order scaling behaviour of $\theta_\pm\left(\omega\right)$ is, in fact, fixed by the dimension of the operator dual to $\varphi_\pm$ in the near-horizon CFT$_1$. At leading order,
\begin{equation}
\theta_+\left(\omega\right)\sim\mathcal{G}_1\left(\omega\right)=i\omega,\;\;\;\;\;\;\;\;\;\;\;\;\;\;\;\;\theta_-\left(\omega\right)\sim\mathcal{G}_2\left(\omega\right)=i\omega^3,
\end{equation}
where $\mathcal{G}_\Delta\left(\omega\right)$ is the retarded Green's function of a scalar operator of dimension $\Delta$ in the near-horizon CFT$_1$. It is now clear that due to the overlap of $J^y$ with the $\Delta=1$ operator in the near-horizon CFT, the conductivity $\sigma\left(\omega\right)$ will have a real part that scales as $\mathcal{G}_1\left(\omega\right)/\omega\sim\omega^0$ as compared to the $m_\beta=0$ case where it has a delta function plus a part that scales as $\sim\mathcal{G}_2\left(\omega\right)/\omega\sim\omega^2$. This is the finite DC conductivity we anticipated due to the non-conservation of momentum.

To see this explicitly, we can determine the retarded Green's function of $J^y$ from (\ref{eq:zeroTonshellactionresult}) via the usual method and then, using the Kubo formula (\ref{eq:kuboformulaconductivity}), the conductivity is
\begin{equation}
\begin{aligned}
\label{eq:ZeroTACConductivityResult}
\sigma\left(\omega\right)&=\frac{L^2}{2\kappa_4^2}\frac{\left(6-m_\beta^2r_0^2\right)}{m_\beta^2r_0^2}\frac{1+\ldots}{1+\kappa_1i\omega\log\left(\frac{\omega r_0}{6-m_\beta^2r_0^2}\right)+\kappa_2\omega+\kappa_3i\omega+\ldots}\\
&=\frac{\sigma_{\text{DC}}+\ldots}{1+\kappa_1i\omega\log\left(\frac{\omega r_0}{6-m_\beta^2r_0^2}\right)+\kappa_2\omega+\kappa_3i\omega+\ldots},
\end{aligned}
\end{equation}
where the ellipses denote terms of order $\omega^2$ and higher, the constants $\kappa_i$ were given previously in (\ref{eq:kappadefinitions}), and the DC conductivity of the state is given by
\begin{equation}
\label{eq:OurDCConductivityResult}
\sigma_\text{DC}\equiv\sigma\left(\omega=0\right)=\frac{L^2}{2\kappa_4^2}\frac{\left(6-m_\beta^2r_0^2\right)}{m_\beta^2r_0^2}.
\end{equation}
The first thing to note is that in the massless limit $m_\beta\rightarrow0$, $\sigma_\text{DC}$ diverges. This divergence indicates the formation of a delta function $\delta\left(\omega\right)$ in the conductivity when momentum is conserved, as we expected.

If we expand in the small mass limit $m_\beta^2r_0^2\ll\omega r_0\ll1$, the presence of the $\sim(m_\beta r_0)^{-2}$ term in $\kappa_3$ leads to
\begin{equation}
\label{eq:ApproxofAcConductivitytoDrudeModel}
\sigma\left(\omega\right)=\frac{\sigma_\text{DC}}{1-i\omega\tau_\text{rel.}},
\end{equation}
where
\begin{equation}
\label{eq:zeroTsmallmassrelaxationtime}
\tau_\text{rel.}=\frac{6}{m_\beta^2r_0}.
\end{equation}
This is precisely the Drude model result! This should not be too surprising, as the small mass limit is the limit where momentum conservation is only weakly violated (or equivalently, where the momentum relaxation time $\mu\tau_\text{rel.}$ is very large). Our full expression (\ref{eq:ZeroTACConductivityResult}) for the conductivity therefore incorporates both the Drude model, and the leading corrections to it. We will shortly explore these corrections in more detail, although unfortunately we do not have any microscopic field theoretic explanation of them.

We note that the momentum relaxation time $\tau_\text{rel.}$ for small masses (\ref{eq:zeroTsmallmassrelaxationtime})  -- i.e. the relaxation time in the limit $\mu\tau_\text{rel.}\gg1$ -- coincides with the naive $T=0$ limit of the hydrodynamic momentum relaxation time (\ref{eq:ourvalueoftau}). We expect that, as for the shear diffusion constant in the massless theory \cite{Davison:2013bxa}, the limit of applicability of the hydrodynamic result (\ref{eq:ourvalueoftau}) is in fact $\tau_\text{rel.}\left(T+\mu\right)\gg1$. As in the hydrodynamic limit, the system absorbs momentum and is therefore unstable unless $\tau_\text{rel.}\geq0$ i.e. $m_\beta^2\geq0$.

Finally, we note that the DC resistivity $\rho_\text{DC}$ is non-zero at $T=0$. This was also noted by taking the low temperature limit in \cite{Vegh:2013sk}. It was shown in \cite{Hartnoll:2012rj} (and confirmed in \cite{Horowitz:2012ky}) that for a theory with a near-horizon AdS$_2$ geometry that is coupled to a lattice with wavevector $k_L$, the DC resistivity has a power law dependence of the form $\rho_\text{DC}\sim T^{2\nu\left(k_L\right)-1}$ for a given function $\nu\left(k_L\right)$. In the limit $k_L\rightarrow0$, this becomes $\rho_\text{DC}\sim T^0$ which is what we have found here, although the prefactor of the power law $T$-dependence in $\rho_{\text{DC}}$ vanishes as $k_L\rightarrow0$. In \cite{Anantua:2012nj}, it was shown for a class of holographic models which have a near-horizon geometry conformal to AdS$_2$ that the DC resistivity due to scattering from random impurities is of the form $\rho_{\text{DC}}\sim T^{2\nu\left(0\right)-1}$ (although things are more complicated when the near-horizon geometry is AdS$_2$ itself \cite{Hartnoll:2012rj}). It would be interesting to further explore this potential connection with massive gravity.

\subsection{Corrections to the Drude model, and the scaling region}

\begin{figure*}[t]
\begin{center}
\includegraphics[scale=0.89]{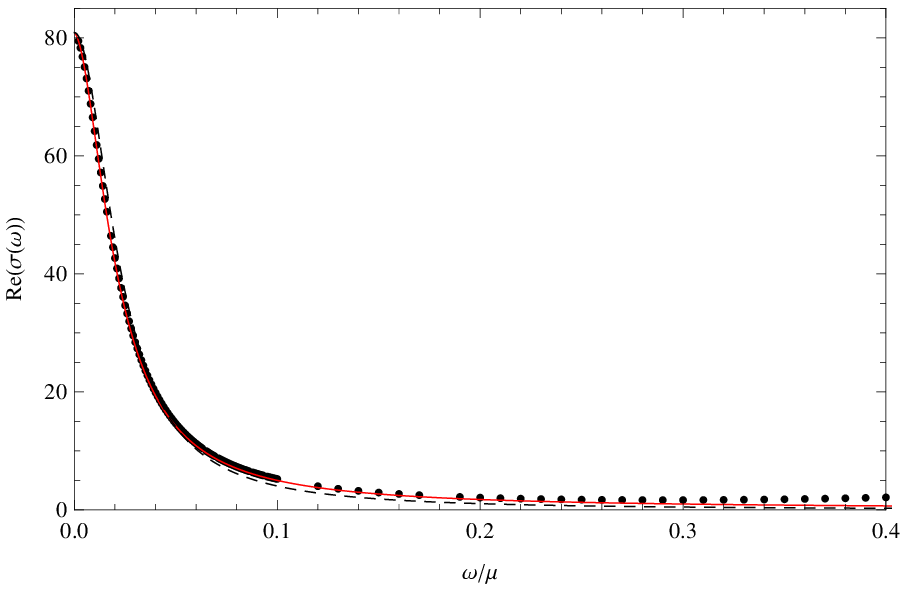}
\includegraphics[scale=0.89]{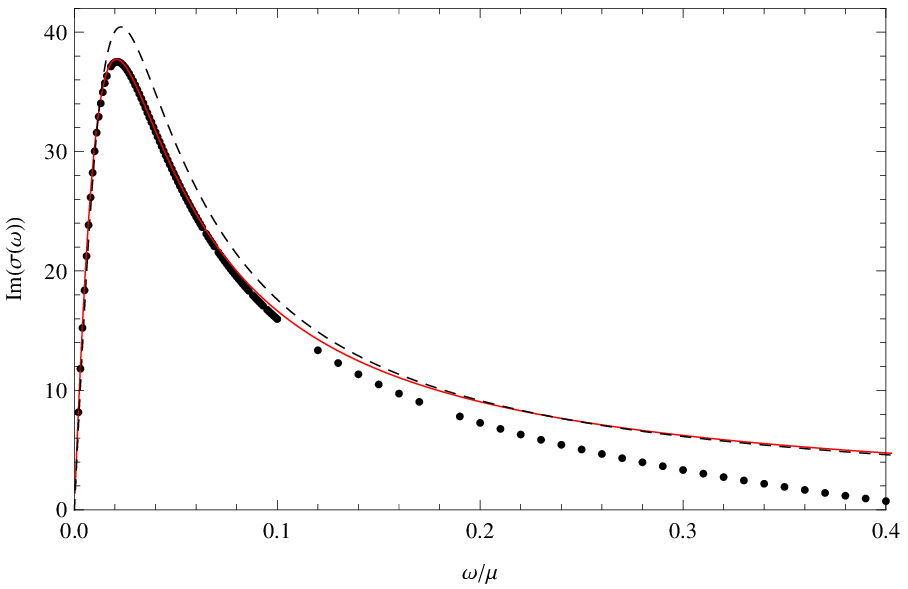}
\includegraphics[scale=0.89]{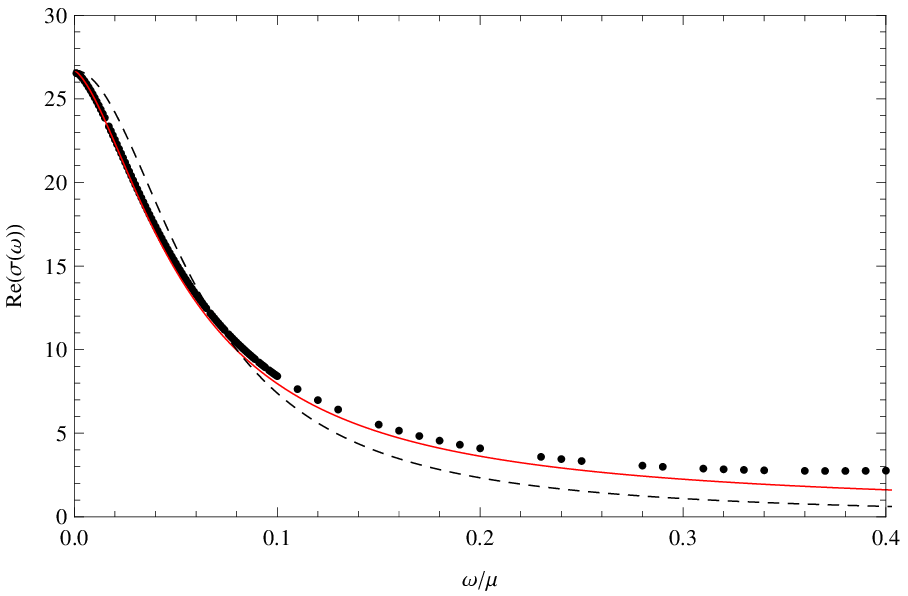}
\includegraphics[scale=0.89]{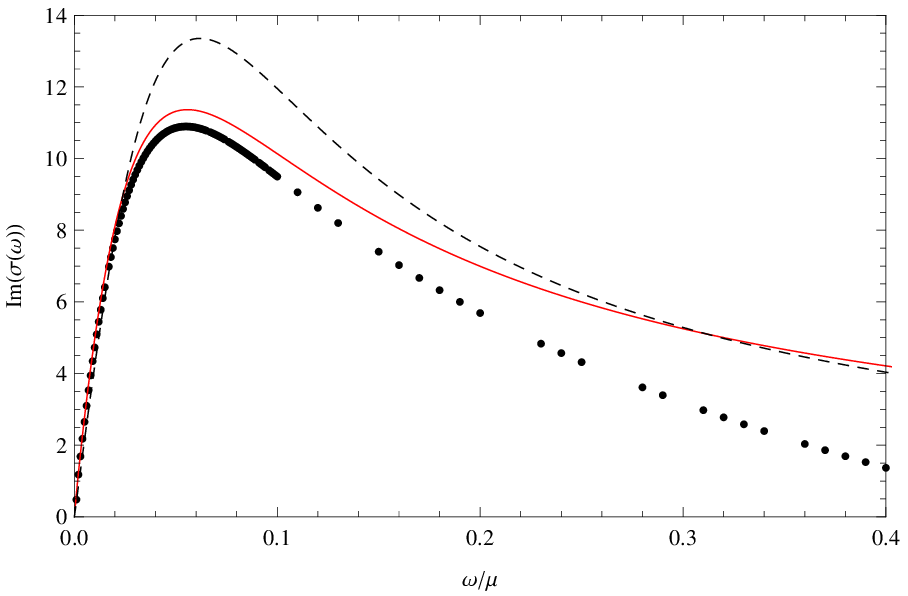}
\caption{The real and imaginary parts of the conductivity $\sigma$ for $m_\beta^2r_0^2\approx0.44$ (top) and $m_\beta^2r_0^2\approx1.04$ (bottom). The black dots are the exact numerical results, the dashed black line is the Drude conductivity (\ref{eq:ApproxofAcConductivitytoDrudeModel}) and the solid red line is the result (\ref{eq:ZeroTACConductivityResult}) which includes corrections to the simple Drude formula. $\sigma$ is plotted in units of $L^2/\left(4\kappa_4^2r_0\mu\right)$.}
\label{fig:ConductivityForVariousMasses}
\end{center}
\end{figure*}

To illustrate the effect of the non-Drude corrections to the conductivity, we show in figure \ref{fig:ConductivityForVariousMasses} the exact conductivity $\sigma\left(\omega\right)$ for two different values of $m_\beta^2r_0^2$, as computed numerically by integrating the equations of motion, and both the Drude conductivity (\ref{eq:ApproxofAcConductivitytoDrudeModel}) and the more accurate result (\ref{eq:ZeroTACConductivityResult}) which includes corrections to the Drude formula.
This figure illustrates a few important points. Firstly, at small $\omega$, the expression (\ref{eq:ZeroTACConductivityResult}) gives a good approximation to the exact conductivity. For small values of $m_\beta^2r_0^2$, there is a well-defined Drude-like peak near the origin of $\text{Re}\left(\sigma\right)$ and as $m_\beta^2r_0^2$ increases, this peak spreads out, transferring its spectral weight to higher $\omega$. In the extreme limit $m_\beta^2r_0^2=3$ (which corresponds to $r_0\mu=0$), the peak completely disappears and the low energy $\text{Re}\left(\sigma\right)$ is constant. This is simply because the current decouples from the momentum in this limit and so can relax quickly. The range of $\omega$ over which our result (\ref{eq:ZeroTACConductivityResult}) is valid decreases as $r_0^2m_\beta^2$ increases (i.e. as the Drude peak spreads out more). This is expected because our result is an expansion in powers of small $\omega r_0/\left(6-m_\beta^2r_0^2\right)^2$ and therefore as $m_\beta^2r_0^2$ increases, our result (\ref{eq:ZeroTACConductivityResult}) is valid over a smaller range of frequencies. 

Secondly, the consequences of the Drude peak spreading out as $m^2_\beta r_0^2$ increases are that $\text{Re}\left(\sigma\right)$ is enchanced at larger $\omega$ while $\text{Im}\left(\sigma\right)$ is suppressed, relative to the Drude result. The corrections to the Drude model encapsulated in (\ref{eq:ZeroTACConductivityResult}) capture these deviations to some extent, but not completely. For the Drude model at large frequencies $\omega\tau_\text{rel.}\gg1$, the phase of the conductivity $\arg\sigma=90^\circ$. The corrections to the Drude model just described reduce this phase, with the amount of reduction increasing with $r_0^2m_\beta^2$.

The phase is a particularly interesting quantity because certain high-$T_c$ superconductors in the normal phase have been found to exhibit scaling law conductivities of the form \cite{experimentalpaper1}
\begin{equation}
\label{eq:ExperimentalConductivityScalingResult}
\sigma\left(\omega\right)=K\omega^{\gamma-2}e^{i\frac{\pi}{2}\left(2-\gamma\right)},
\end{equation}
with $\gamma\approx1.35$ and a constant $K$, in the range of frequencies corresponding to $\omega\gg T,\tau_\text{rel.}^{-1}$ (up to some high energy cutoff which for us is $\sim\mu$). This scaling law, which is taken as a signal of underlying quantum criticality and is currently unexplained, manifests itself in two clearly observable ways: scaling behaviour of the magnitude of the conductivity $\left|\sigma\right|=K\omega^{\gamma-2}$ and a constant phase $\arg\sigma=\left(2-\gamma\right)90^\circ$.

\begin{figure*}
\begin{center}
\includegraphics[scale=0.92]{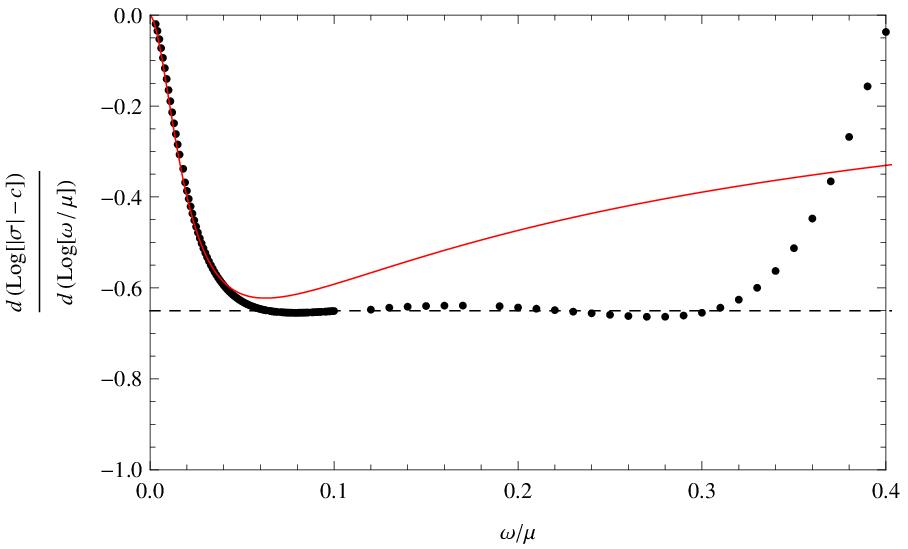}
\includegraphics[scale=0.86]{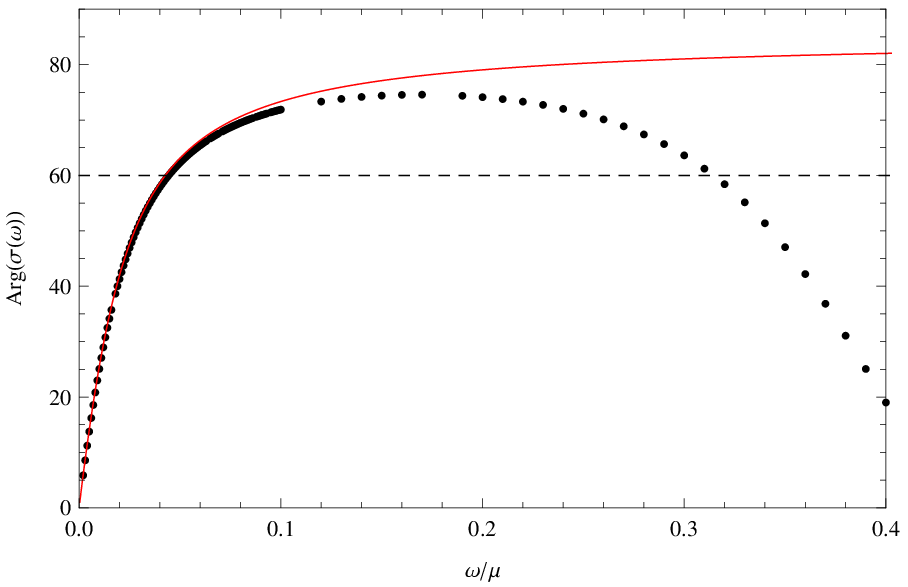}
\caption{The potential scaling exponent (left) as described in the main text with $c=-8.9$, and phase (right), of the conductivity $\sigma$ when $m_\beta^2r_0^2\approx0.44$. Black dots show the exact numerical results, the solid red line is the analytic expression (\ref{eq:ZeroTACConductivityResult}) and the dashed black lines denote the exponent and phase found experimentally in \cite{experimentalpaper1}. Both $c$ and $\sigma$ are given in units of $L^2/\left(4\kappa_4^2r_0\mu\right)$.}
\label{fig:ConductivityArgumentAndExponent}
\end{center}
\end{figure*}

In \cite{Vegh:2013sk}, it was shown that at small temperatures the conductivity of the theory dual to (\ref{eq:MassiveGravityAction}) behaves like a scaling law with a constant term $\left|\sigma\right|=c+K\omega^{\gamma-2}$, and an approximately constant phase. Similar behaviour with a robust value of $\gamma\approx1.35$ was seen in the holographic lattice models \cite{Horowitz:2012ky,Horowitz:2012gs}. In figure \ref{fig:ConductivityArgumentAndExponent} we plot $d\log\left(\left|\sigma\right|-c\right)/d\log\omega$ (which should be a constant equal to $\gamma-2$ if $\sigma$ is described by the expression above) and the phase of $\sigma$. 
By increasing $r_0m_\beta$ from zero to its maximum value, one can obtain values of $\gamma$ between $1$ and $2$, and phases between $90^\circ$ and $0^\circ$. We have chosen $r_0^2m_\beta^2\approx0.44$ such that $\gamma$, as extracted from the logarithmic plot, is $\approx1.35$. The phase is not constant, but is relatively close to $\left(2-\gamma\right)90^\circ\approx60^\circ$ over a large range of $\omega$.

One goal of our analytic computation of the low $\omega$ conductivity was to better understand these results. Specifically, it was to determine whether there really is a scaling behaviour similar to (\ref{eq:ExperimentalConductivityScalingResult}) (with an extra constant term), and whether this can be understood as arising from the semi-local quantum critical state dual to the near-horizon AdS$_2$ geometry. Although our analytic result (\ref{eq:ZeroTACConductivityResult}) does not show any sign of this scaling behaviour, it can be seen in figure \ref{fig:ConductivityArgumentAndExponent} that our analytic results are unfortunately not close enough to the exact result in the regime of interest to conclusively determine whether there is such a power law or not. However, we assert that any potential scaling behaviour of this type in the conductivity cannot be simply traced back to the presence of a near-horizon AdS$_2$ geometry, as our calculation captures accurately the effects of this region of the spacetime upon $\sigma\left(\omega\right)$, and does not produce a power law. Interpreting the radial direction as an energy scale in the usual way, it seems that the region of $\omega$ over which there is potentially a scaling law is far enough from $\omega=0$ that the observables in this region should be strongly affected by regions of $r$ which are not infinitesimally close to $r_0$, i.e. by the geometry not next to the horizon. For this reason, it would be interesting to study geometries in which the spacetime has a scaling symmetry over a range of intermediate values of $r$ between the horizon and boundary. Similar conclusions were reached for models incorporating an explicit bulk lattice in \cite{Horowitz:2013jaa}.

\section{A preliminary search for spatially modulated instabilities}
\label{sec:PreliminaryStabilityStudy}

As we have emphasised, the equilibrium properties of the field theory state we are studying exhibit spatial translational invariance. One may worry that the preferred equilibrium solution may be a state without such a symmetry. In this section we will search for a specific kind of instability of the solution (\ref{eq:backgroundsolution}) with $m_\alpha=0$, which would result in the formation of spatially modulated expectation values for $J^y$, $T^{ty}$ or $T^{yx}$ in the dual field theory, and find that no such instability occurs.

The existence of instabilities leading to spatially modulated equilibrium states are well-known in holographic models of states of matter at finite density. One source of such an instability is when the momentum-dependent mass of a field violates the Breitenlohner-Freedman (BF) bound \cite{Breitenlohner:1982bm} of the near-horizon geometry over a range of non-zero momenta \cite{Nakamura:2009tf,Donos:2011bh,Donos:2011ff,Donos:2012ra}. This type of instability is often associated with Chern-Simons or axionic terms in the bulk action, neither of which are present here, but it was recently shown in \cite{Donos:2013gda} that instabilities of this kind are also possible when such terms are not present. The violation of the near-horizon BF bound is a well-studied topic in holography \cite{Kaplan:2009kr,Jensen:2010ga,Iqbal:2010eh,Jensen:2010vx,Pal:2010gj,Iqbal:2011ae} and typically leads to a BKT-type transition in the dual field theory.

It is relatively straightforward to identify any such spatially modulated instabilities in our theory. We will work at $T=0$ and examine the masses of the transverse fields that we have studied throughout this work. As in the previous section, for simplicity we will restrict ourselves to the case $m_\alpha=0$, where we may write the equations of motion in the relatively simple form (\ref{eq:VarPhi1EoM}), (\ref{eq:VarPhi2EoM}) and (\ref{eq:VarPhi3EoM}) for the three independent fields $\varphi_{1,2,3}$ given in (\ref{eq:defnofvarphiifields}). To access the near-horizon AdS$_2$ region, we change co-ordinates to (\ref{eq:ZetaCoordinateDefinition}) and expand the resulting equations of motion as power series in $\omega$. At lowest order in $\omega$, the equations of motion are
\begin{subequations}
\begin{align}
& \varphi_1''+\varphi_1-\frac{r_0^2\left(2m_\beta^2+k^2\right)}{\left(6-r_0^2m_\beta^2\right)\zeta^2}\varphi_1-\frac{2\sqrt{3-r_0^2m_\beta^2}\left(2m_\beta^2+k^2\right)}{\left(6-r_0^2m_\beta^2\right)\zeta^2}\varphi_2=0, \label{eq:NearHorizonFiniteKEq1}\\
& \varphi_2''+\varphi_2-\frac{r_0^2k^2+4\left(3-r_0^2m_\beta^2\right)}{\left(6-r_0^2m_\beta^2\right)\zeta^2}\varphi_2-\frac{2r_0^2\sqrt{3-r_0^2m_\beta^2}}{\left(6-r_0^2m_\beta^2\right)\zeta^2}\varphi_1=0, \label{eq:NearHorizonFiniteKEq2}\\
& \varphi_3''+\varphi_3=0. \label{eq:NearHorizonFiniteKEq3}
\end{align}
\end{subequations}
These equations may be diagonalised by changing variables to
\begin{equation}
\begin{aligned}
&\phi_1=\varphi_1+\frac{1}{2r_0^2\sqrt{3-r_0^2m_\beta^2}}\left[3\left(2-r_0^2m_\beta^2\right)+\sqrt{\left(6-r_0^2m_\beta^2\right)^2+4k^2r_0^2\left(3-r_0^2m_\beta^2\right)}\right]\varphi_2,\\
&\phi_2=\varphi_1+\frac{1}{2r_0^2\sqrt{3-r_0^2m_\beta^2}}\left[3\left(2-r_0^2m_\beta^2\right)-\sqrt{\left(6-r_0^2m_\beta^2\right)^2+4k^2r_0^2\left(3-r_0^2m_\beta^2\right)}\right]\varphi_2,\\
&\phi_3=\varphi_3,
\end{aligned}
\end{equation}
whose equations of motion are simply the equations of motion of scalar fields in AdS$_2$ (after rescaling $\zeta$ by a factor of $\omega$) with masses
\begin{equation}
\begin{aligned}
&\left(m_1L_2\right)^2=\frac{1}{\left(6-r_0^2m_\beta^2\right)}\left[6-r_0^2m_\beta^2+k^2r_0^2+\sqrt{\left(6-r_0^2m_\beta^2\right)^2+4k^2r_0^2\left(3-r_0^2m_\beta^2\right)}\right],\\
&\left(m_2L_2\right)^2=\frac{1}{\left(6-r_0^2m_\beta^2\right)}\left[6-r_0^2m_\beta^2+k^2r_0^2-\sqrt{\left(6-r_0^2m_\beta^2\right)^2+4k^2r_0^2\left(3-r_0^2m_\beta^2\right)}\right],\\
&\left(m_3L_2\right)^2=0.
\end{aligned}
\end{equation}
For real $k$, and $0\leq m_\beta^2r_0^2\leq3$ (the range of values of $m_\beta^2r_0^2$ where the previously-identified $k=0$ instability does not occur), it is clear that $\left(m_1L_2\right)^2,\left(m_3L_2\right)^2\ge0$. The BF bound in AdS$_2$ is $\left(mL_2\right)^2>-1/4$ and so there is no instability due to $\phi_{1,3}$. For $\phi_2$, this is not so obvious. If we consider a fixed bulk graviton mass $m_\beta$, $\left(m_2L_2\right)^2\left(k\right)$ is an even function of $k$ and goes from $0$ at $k=0$ to $+\infty$ as $k\rightarrow\infty$. By taking a partial derivative with respect to $k$, its turning points $k_*$ are given by
\begin{equation}
k_*^2r_0^2=0,\;\;\;\;\;\;\;\;\;\;\;\;\;\;\;\text{and}\;\;\;\;\;\;\;\;\;\;\;\;\;\;\;k_*^2r_0^2=-\frac{3r_0^2m_\beta^2}{4}\frac{\left(4-r_0^2m_\beta^2\right)}{\left(3-r_0^2m_\beta^2\right)}.
\end{equation}
For real $k$, and $0\leq m_\beta^2r_0^2<3$, this second equation has no solutions and thus the only turning point is at $k=0$. In other words, $\left(m_2L_2\right)^2$ is an even function of $k$ which monotonically increases from $0$ to $\infty$ as $k$ goes from $0$ to $\infty$. Therefore it never violates the AdS$_2$ BF bound.

A real lattice of course has spatial modulation of equilibrium observables (such as charge density and energy density) corresponding to longitudinal fields, and it is in the longitudinal sector that the type of non-zero $k$ instability discussed here is present in \cite{Donos:2013gda}. It is therefore clearly important to repeat the above calculation for the longitudinal fields, but this is beyond the scope of this paper.

We emphasise here that although we have ruled out instabilities of the nature just described, there are of course other possible sources of $k\ne0$ instabilities in the transverse fields that we have not investigated -- for example, those associated with the dynamics of the fields outside the near-horizon region and also those of states with $m_\alpha\ne0$ -- as well as instabilities in the longitudinal fields as mentioned above. To identify instabilities arising outside the near-horizon region, one should compute the quasinormal mode spectrum and determine if there are any modes with positive imaginary part (e.g. by applying Leavers method \cite{Leaver:1990zz} to the equations of motion (\ref{eq:VarPhi1EoM}), (\ref{eq:VarPhi2EoM}) and (\ref{eq:VarPhi3EoM})). There is also the possibility of the existence of first order transitions to more stable states, which will not be identified even by this quasinormal mode analysis.

\section{Discussion}
\label{sec:DiscussionSection}

In summary, we have investigated the effect of the relaxation of momentum upon various observables in the field theory holographically dual to the massive gravity action (\ref{eq:MassiveGravityAction}). In the hydrodynamic limit, the dominant effect of this relaxation can be incorporated into the low energy effective theory by a modification of the conservation equation of energy-momentum to (\ref{eq:nonconservationproposal}), such that momentum is no longer conserved. We computed the characteristic timescale over which momentum relaxes in this limit (\ref{eq:ourvalueoftau}) and identified the ``wall of stability'' found in \cite{Vegh:2013sk} with the point at which this timescale vanishes. We also computed analytically the AC conductivity of this theory at zero temperature (\ref{eq:ZeroTACConductivityResult}), including corrections to the simple Drude model result. Finally, we made a preliminary study of the possible existence of $k\ne0$ instabilities, and didn't find any due to near-horizon BF bound violation by the transverse fields.

Our results are broadly what one would expect of a physical system which dissipates momentum and further justify the use of the massive gravity action (\ref{eq:MassiveGravityAction}) as a holographic model of such a system. Having calculated the momentum relaxation timescale and conductivity in terms of the graviton masses, we would like to know what microscopic field theoretical quantities (if any) these graviton masses correspond to. To do this, it is clearly important to better understand in what sense we can discuss a possible field theory dual of (\ref{eq:MassiveGravityAction}). For example, can the theory (\ref{eq:MassiveGravityAction}) be derived as the long-distance effective theory of a conventional holographic system (i.e. one with massless gravitons) in which translational invariance is broken by a well-understood microscopic mechanism such as by coupling to impurities \cite{Hartnoll:2007ih,Hartnoll:2008hs} or by the presence of a lattice \cite{Horowitz:2012ky,Horowitz:2012gs,Horowitz:2013jaa,Hartnoll:2012rj}? We note that one can generate a mass term for gravitons in a holographic setup by coupling two CFTs such that their dual theory comprises two AdS spaces with a shared boundary \cite{Kiritsis:2006hy,Aharony:2006hz,Kiritsis:2008xj,Kiritsis:2008at,Kiritsis:2011zq,Apolo:2012gg}. In this setup, the energy-momentum tensor of each individual CFT is not conserved due to interactions with the other.

Additionally, there are a few more straightforward calculations which would be of interest. One is to examine the hydrodynamic excitations in the longitudinal sector to see if they agree with the modified version of hydrodynamics we have proposed, and a second is to perform a more complete stability analysis of the solution along the lines outlined in section \ref{sec:PreliminaryStabilityStudy}, with particular attention paid to the possible existence of spatially inhomogeneous instabilities. 

\subsection*{Acknowledgements}

I would like to thank Sean Hartnoll, Elias Kiritsis, Andrei Parnachev, Koenraad Schalm, David Vegh and Jan Zaanen for helpful discussions and correspondence. I am especially grateful to A. Parnachev and D. Vegh for their comments on a draft of this manuscript. This work was supported by a VIDI innovative research grant from NWO, The Netherlands Organisation for Scientific Research.

\appendix
\section{Appendix: Simplified equations of motion for $m_\alpha=0$}

For solving the equations (\ref{eq:HytEoM}), (\ref{eq:HyxEoM}), (\ref{eq:HyrEoM}) and (\ref{eq:ayEoM}) in the $m_\alpha=0$ limit, it is convenient to use the variables
\begin{equation}
\begin{aligned}
\label{eq:defnofvarphiifields}
\varphi_1=\frac{1}{r^2}\left({h^y_t}'+i\omega h^y_r\right)+A_t'a_y,\;\;\;\;\;\;\;\;\varphi_2=a_y,\;\;\;\;\;\;\;\;\varphi_3=\frac{f}{r^2}\left({h^y_x}'-ikh^y_r\right),
\end{aligned}
\end{equation}
which are linear combinations of the fundamental fields and their derivatives. As was the case when we used the fundamental fields $h^y_t$, $h^y_x$, and $a_y$, there are three linearly-independent dynamical variables $\varphi_i$. In the $m_\alpha=0$ limit, the equations of motion can be written in terms of these variables as
\begin{subequations}
\begin{align}
& \frac{d}{dr}\left[r^2f\varphi_1'\right]-k\left(k{h^y_t}'+\omega{h^y_x}'\right)-2m_\beta^2{h^y_t}'=0,\\
& \frac{d}{dr}\left[f\varphi_2'\right]+r^2A_t'\varphi_1+\left[\frac{1}{f}\left(\omega^2-k^2f\right)-r^2{A_t'}^2\right]\varphi_2=0,\\
& \frac{d}{dr}\left[r^2f\varphi_3'\right]+\omega\left(k{h^y_t}'+\omega{h^y_x}'\right)=0,
\end{align}
\end{subequations}
in addition to the constraint (\ref{eq:HyrEoM}). Using this constraint and the definitions of the fields $\varphi_i$ in (\ref{eq:defnofvarphiifields}), these become
\begin{subequations}
\begin{align}
& \frac{d}{dr}\left[r^2f\varphi_1'\right]+\frac{r^2}{f}\left[\omega^2-k^2f-2fm_\beta^2\right]\varphi_1+r^2{A_t'}\left(2m_\beta^2+k^2\right)\varphi_2=0, \label{eq:VarPhi1EoM}\\
& \frac{d}{dr}\left[f\varphi_2'\right]+r^2A_t'\varphi_1+\left[\frac{1}{f}\left(\omega^2-k^2f\right)-r^2{A_t'}^2\right]\varphi_2=0, \label{eq:VarPhi2EoM}\\
& \frac{d}{dr}\left[r^2f\varphi_3'\right]+\frac{\omega^2r^2}{f}\varphi_3+\omega kr^2\varphi_1-\omega kr^2{A_t'}\varphi_2=0. \label{eq:VarPhi3EoM}
\end{align}
\end{subequations}

\subsection*{Decoupling of fields when $k=0$}

In the $k=0$ limit, the field $\varphi_3$ decouples from $\varphi_1$ and $\varphi_2$. This is analogous to the decoupling of $h^y_x$ from $h^y_t$ and $a_y$ in the same limit. By defining the field $\tilde{\varphi}_1\equiv r\varphi_1$ and partially expanding the derivative in the $\varphi_1$ equation of motion (\ref{eq:VarPhi1EoM}), we find that the coupled equations are
\begin{subequations}
\begin{align}
& \frac{d}{dr}\left[f\tilde{\varphi}_1'\right]+\left(\frac{\omega^2}{f}-2m_\beta^2-\frac{f'}{r}\right)\tilde{\varphi}_1+2m_\beta^2rA_t'\varphi_2=0, \label{eq:precombinationeq1}\\
& \frac{d}{dr}\left[f\varphi_2'\right]+\left(\frac{\omega^2}{f}-r^2{A_t'}^2\right)\varphi_2+rA_t'\tilde{\varphi_1}=0. \label{eq:precombinationeq2}
\end{align}
\end{subequations}
Taking the linear combination $\gamma\times(\ref{eq:precombinationeq1})+(\ref{eq:precombinationeq2})$, we find that the equations of motion take the form
\begin{equation}
\label{eq:almostdecoupledeqns}
\frac{d}{dr}\left[f\varphi_2'\right]+\gamma\frac{d}{dr}\left[f\tilde{\varphi}_1'\right]+\left(\frac{\omega^2}{f}-r^2{A_t'}^2+2\gamma m_\beta^2rA_t'\right)\left(\varphi_2+\gamma\tilde{\varphi}_1\right)=0,
\end{equation}
provided that $\gamma$ satisfies the quadratic equation
\begin{equation}
-2m_\beta^2 rA_t'\gamma^2+\gamma\left(r^2{A_t'}^2-2m_\beta^2-\frac{f'}{r}\right)+rA_t'=0.
\end{equation}
The solutions to this quadratic equation, given by
\begin{equation}
\gamma_\pm=-\frac{3}{4m_\beta^2r_0^2\mu}\left[\left(1-r_0^2m_\beta^2+\frac{1}{4}r_0^2\mu^2\right)\pm\sqrt{\left(1-r_0^2m_\beta^2+\frac{1}{4}r_0^2\mu^2\right)^2+\frac{8}{9}m_\beta^2\mu^2r_0^4}\right],
\end{equation}
are in fact independent of $r$, and thus we can write the equations of motion (\ref{eq:almostdecoupledeqns}) in the decoupled form (\ref{eq:nomomentumequations}) for the variables (\ref{eq:nomomentummasterfields}). These are somewhat analogous to the decoupled Kodama-Ishibashi variables for the massless theory with $k\ne0$ \cite{Kodama:2003kk}.

\singlespace
\bibliographystyle{JHEP}
\bibliography{MassiveGravityv2}

\end{document}